\documentclass[journal,onecolumn,draftcls,12pt]{IEEEtran}
\usepackage{url}
\usepackage{ifthen, color}
\usepackage{cite}
\usepackage{amsmath,amsfonts}
\usepackage{algorithmic}
\usepackage{algorithm}
\usepackage{array}
\usepackage{caption}
\usepackage{subcaption}
\usepackage{makecell}
\usepackage{multirow}
\usepackage{array}
\usepackage{amssymb,graphicx}
\newtheorem{Remark}{Remark}
\usepackage{arydshln}
\allowdisplaybreaks

\IEEEoverridecommandlockouts

\title{Characterizing the Optimal Memory-Rate Tradeoff in Secure Coded Caching for Small Buffer or Small Rate}

\author{
	\IEEEauthorblockN{
		Han Fang, Nan Liu and Wei Kang}	
	\thanks{This paper was presented in part at the 2024 IEEE International Symposium on Information Theory  (ISIT). 
		
		H. Fang and W. Kang are with the School of Information Science and Engineering,
		Southeast University, Nanjing, China (email: \{fanghan, wkang\}@seu.edu.cn). N. Liu is with the National Mobile Communications Research Laboratory,
		Southeast University, Nanjing, China (email:  nanliu@seu.edu.cn). 
		}}

\begin{document}
\maketitle
\begin{abstract}
We consider the secure coded caching problem proposed by Ravindrakumar \emph{et. al} where no  user can obtain information about files other than the one requested. We first propose three new schemes for the three cases of cache size \( M=1\), $N=2$ files and arbitrary $K$ users, delivery rate \( R=1, \) arbitrary $N$ files and $K$ users, and the general case for arbitrary $N$ files and $K$ users, respectively. Then we derive  converse results by characterizing new properties of secure coded caching schemes. As a result, we characterize the two end-points of the optimal memory-rate tradeoff curve for arbitrary number of users and files. Furthermore, for the case of $N=2$ files and arbitrary number of users, we also characterize a segment of the optimal memory-rate tradeoff curve, where the cache size is relatively small.
\end{abstract}
\section{Introduction}
In their seminal work, Maddah-Ali and Niesen \cite{MAN2014} introduced the problem of coded caching,  where there is a server in possession of $N$ files, and $K$ users, each with a cache large enough to store $M$ files. In the cache placement phase, the server fills each user's cache with some file-related contents. In the delivery phase, each user makes a file request, and the server sends a delivery signal through a noiseless broadcast channel to the users to ensure that each user can decode the requested file based on the delivery signal and local cache content. The design aim is to minimize the maximum delivery rate among all requests, also known as the worst-case delivery rate $R$. The goal is to find the optimal memory-rate tradeoff, i.e., the minimum worst-case delivery rate $R$ as a function of the cache size $M$. 
Reference \cite{MAN2014} showed that global caching gain can be achieved by exploiting coded multicast deliveries.

In this paper, we focus on the \textit{secure} coded caching problem, which was first proposed by Ravindrakumar \emph{et. al}\cite{PCC2018}. Secure coded caching adds another constraint to the original coded caching problem, i.e., no user can obtain any information about the files other than the one requested. The secure coded caching problem was further studied in \cite{PDAsecretive2021} and \cite{commonsecretive2017}. More specifically, reference \cite{PDAsecretive2021} shows that the use of placement delivery array constructions can reduce the file subpacketization levels in the secure coded caching problem, and reference\cite{commonsecretive2017} designs new schemes that can achieve smaller \emph{average} rates by considering the fact that different users may request the same file. Furthermore, variations of the secure coded caching problem has been studied in 
 combination networks\cite{Combinationsecure2018, cachingdeliverysecure2019}, D2D networks\cite{D2Dsecure2020}, resolvable networks\cite{resolvablesecure2016}, fog radio access networks\cite{Fogsecretive2023,Jiangyx}, and shared cache structure\cite{ Sharesecretive2021,SharePDAsecretive2022}.  In addition, the secure coded caching problem for colluding users is studied in \cite{colludingsecure2021}, and references \cite{ FDsecure2022,Qichao} consider the security condition together with demand privacy, where users can neither obtain information about unrequested files nor learn about other users' requests.

 In this paper, we follow the model in \cite{PCC2018} and study the secure coded caching problem considering the worst-case delivery rate. It has been shown that for the secure coded caching problem to be feasible, we must have $M \geq 1$ and $R \geq 1$ \cite{PCC2018}. Furthermore, some achievability and converse results are given in \cite{PCC2018} and the ratio of the  achievability and converse results is proved to be constrained within a constant range, which is also known as order-optimal.  The optimal performance, i.e., the optimal memory-rate tradeoff, is still an open problem except for the 2-user 2-file case, i.e., $N=K=2$ \cite{PCC2018}.

 In this paper, we first propose a new achievable scheme for each of the three cases: 1) the smallest cache size for the case of two files and arbitrary number of users, i.e., \( M=1 \), $N=2$ and arbitrary $K$; 2) the smallest delivery rate for arbitrary number of files and users, i.e., \( R=1 \) and arbitrary $N,K$; and 3) the general case, 
 i.e., \( M > 1 \), \( R > 1 \) and arbitrary $N,K$. These schemes achieve a smaller delivery rate for the same cache size compared to  the best known existing results\cite{PCC2018}. Furthermore, by deriving matching converse results,
 we prove that our schemes are optimal 
for the following cases: 1)  \( M=1 \), $N=2$ and arbitrary $K$; 2) \( R=1 \), and arbitrary $N,K$; and 3) \(M \in \left[1, \frac{K}{K-1}\right]\), \(N = 2\), and arbitrary $K$. Moreover, our derived converse results prove the optimality of the scheme proposed in \cite{PCC2018} for the case of  \( M=1 \), \( N>2 \), and arbitrary $K$.  Hence,  for arbitrary number of files and users, the optimal performance for the smallest cache size and the smallest delivery rate  is fully characterized. The main contributions of this paper are

\noindent
1) In terms of achievability, we propose three new linear schemes, where two schemes are for the two extreme cases, i.e.,  $(M,N)=(1,2)$, arbitrary $K$ and $R=1$, arbitrary $N,K$, and one scheme is for the general case, i.e., $M>1,R>1$, arbitrary $N,K$. The new schemes perform better than the best known existing results \cite{PCC2018}. More specifically, when $M=1$, the new achievable rate is $R=K-1$ for $N=2$ and arbitrary $K$, which  is 1 smaller than the best known achievable rate $R=K$ for $N=2$ and $K \geq 3$ \cite{PCC2018}. When $R=1$, our achievable scheme  requires a cache size of $M=(N-1)(K-1)$ for arbitrary $N,K$, which is $K-1$ smaller than the best known required cache size $M=N(K-1)$ for $K \geq 3$ and arbitrary $N$ \cite{PCC2018} . When $M>1,R>1$, for the same rate $	R(M(t)) = \frac{K}{t+1}$ with  $t \in\{1, \ldots, K-2\}$, our achievable scheme  requires a cache size of $M(t)=\frac{Nt}{K-t}+1-\frac{1}{\binom{K}{t}-\binom{K-1}{t-1}}$ for $K \geq 3$ and arbitrary $N$, which is $\frac{1}{\binom{K}{t}-\binom{K-1}{t-1}}$ smaller than the best known cache size $M(t)=\frac{Nt}{K-t}+1$ for $K \geq 3$ and arbitrary $N$ \cite{PCC2018}.

\noindent
2) In terms of converse, for the two extreme cases i.e., $M=1$ and $R=1$, we first characterize some properties that any caching and delivery scheme for the secure coded caching problem must satisfy. More specifically, when $M=1$, we show that the cached content at a user is a deterministic function of the delivery signal and its requested message. This result further leads to the mutual independence of a single file, and the cached content at each user. When $R=1$, we show that the delivery signal is a deterministic function of the cached content of a user and its requested message. This result further leads to the mutual independence of a single file, and some of the delivery signals. Based on these properties, the converse can be derived. Furthermore, for \(N = 2\) and \( K \geq 3\), when \(M \in \left[1, \frac{K}{K-1}\right]\), we utilize  the symmetry property \cite{tianchao} to derive a new converse result, i.e., $(K-1)(K-2)M+2R\ge K(K-1)$.  The new converse is tighter than existing converse results. More specifically, when $M=1$,  the new converse is
\begin{equation*}
{R} \geq 
\begin{cases}
K-1, &{\text{if}}\ N=2,\\
K, &{\text{if}}\ N \geq 3,
\end{cases}
\end{equation*}
which is tighter than the best existing converse $R \geq \min \{N/2,K\}$\cite{PCC2018}. When $R=1$, the cache size must satisfy $M\ge(N-1)(K-1)$ for arbitrary $N,K$, which is a  significantly tighter lower bound than the best existing lower bound of    $M\ge{\lfloor{N/2}\rfloor}$\cite{PCC2018} for $K \geq 3$ and arbitrary $N$. When \(M \in \left[1, \frac{K}{K-1}\right]\) for \(N = 2\) and \( K \geq 3\), the new converse is $(K-1)(K-2)M+2R\ge K(K-1)$ which is tighter than the best existing converse $R \geq 1$\cite{PCC2018}.

\noindent
3) Finally, by comparing our achievability results, our converse results, and existing achievability results, we characterize 1) the two end-points of the optimal  memory-rate tradeoff curve for arbitrary number of users and files; and 2) the optimal memory-rate tradeoff curve for a segment where the cache size is relatively small in the case of two files and arbitrary number of users. 

The rest of the paper is organized as follows. We introduce the system model of secure coded caching in Section \ref{section2} and present the main results of this paper in Section \ref{section3}. The achievability proofs of our main results are given  in Section \ref{section4} and the converse proofs are given in Section \ref{section5}. At last, we conclude this paper in Section \ref{section6}.

\section{System Model}\label{section2}

\textit{Notation}: We use capital letters to represent random variables, capital letters in calligraphic to represent sets, and letters in bold to represent vectors and matrices. We use $\mathbb{F}_{q}$  to denote the finite field of size  $q$  where  $q$  is a prime power. For a positive integer  $n$, $\mathbb{F}_{q}^{n}$  is the  $n$-dimensional vector space over $\mathbb{F}_{q}$. To simplify the notation, we denote the set $\{1,2, \ldots, n\}$ as $[n]$ and $\{n_1, n_1+1, \cdots, n_2\}$ as $[n_1:n_2]$. For an index set $\mathcal{A}=\{a_t|t\in [n]\}$, denote the vector $(Y_{a_1},Y_{a_2},\ldots,Y_{a_n})$ as $\mathbf{Y}_\mathcal{A}$. Let $\mathbf{e}_i$ denote a column vector where only the $i$-th element is $1$ and the other elements are 0. $\mathbf{I}$ represents the identity matrix.  The dimensions of the vectors and matrices will be clear from the context. We use $r(\mathbf{G})$ to represent the rank of the matrix $\mathbf{G}$. We use the notation $\oplus_q$ and $\ominus_q$ to represent symbol-wise addition and subtraction in finite field $\mathbb{F}_{q}$ respectively, and we may omit $q$ when there is no ambiguity.

\begin{figure}[t]
	\centering
	\includegraphics[width=.55\textwidth]{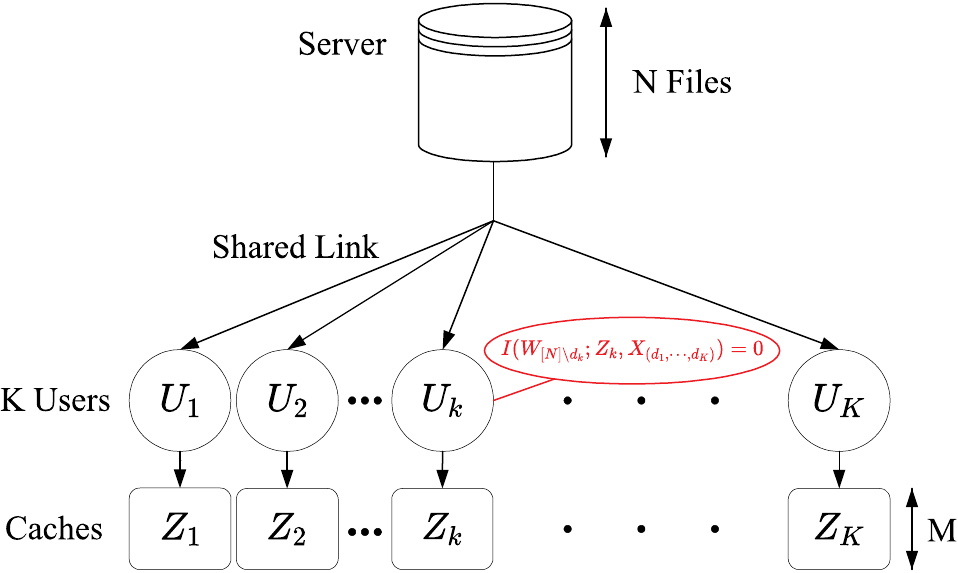}
	\caption{System model.}
	\label{fig: figure1}
\end{figure}

In this paper, we study the secure coded caching problem proposed by Ravindrakumar \emph{et. al} \cite{PCC2018} shown in Fig. \ref{fig: figure1}. The system consists of a server storing $N$ files, and $K$ users, who can receive information through a noiseless broadcast channel from the server. Assume that the $N$ files are independent, and each consists of $F$ i.i.d. symbols uniformly distributed on $\mathbb{F}_{q}$. The files are denoted as $\mathbf{W}=\left(W_{1}, W_{2}, \ldots, W_{N}\right)$. The users each has a cache large enough to store $M$ files, where $M$  denotes the \emph{cache size}. To achieve secure delivery, some randomness is generated at the server. We define the randomness as a random variable $S$, which is independent to the files and consists of $LF$ i.i.d. symbols uniformly distributed on $\mathbb{F}_{q}$, where $L$ is called the \emph{amount of randomness}.

The process includes two phases: the \textit{placement phase} and the \textit{delivery phase}. During the placement phase,  user $k$ receives content $Z_k$ from the server and places it in its cache 
\begin{align}
	Z_{k}\mathrel{\triangleq}\phi_{k}(W_{1},{\ldots}W_{N},S),\nonumber
\end{align}
where $\phi_{k}\colon \mathbb{F}_{q}^{NF}\times \mathbb{F}_{q}^{LF}\to \mathbb{F}_{q}^{{\lfloor{MF}\rfloor}}$ is the $k$-th caching function. 
Note that the size of the content can not exceed the cache size of the users, i.e., $MF$. 
	
During the delivery phase, each user $k$ requests one of the $N$ files, and the index of the requested file is denoted as $d_k$. The server delivers a signal $	X_{(d_{1},\ldots, d_{K})}$ of size $RF$ to all the users over the noiseless broadcast channel, i.e., 
\begin{align}
	X_{(d_{1},\ldots, d_{K})}\mathrel{\triangleq}\psi_{(d_{1},\ldots, d_{K})}(W_{1},\ldots, W_{N},S),\nonumber
\end{align}
where $\psi_{(d_{1},{\ldots}, d_{K})}\colon$  $\mathbb{F}_{q}^{NF}\times \mathbb{F}_{q}^{LF}\to \mathbb{F}_{q}^{{\lfloor{RF}\rfloor}}$ is the encoding function, and $R$ is called the \emph{worst-case delivery rate}. We emphasize here that the placement phase happens before the users make requests for the files. Thus, the design of cache contents $Z_1, \cdots, Z_k$ can not depend on the knowledge of the user demands $(d_{1},\ldots, d_{K})$. 

Upon receiving the delivery signal $X_{(d_{1},\ldots, d_{K})}$, each user $k$ recovers the requested file $W_{d_k}$ from $X_{(d_{1},\ldots, d_{K})}$ transmitted by the server and its own local cache content $Z_k$. For a secure coded caching scheme, two conditions need to be satisfied, namely, the correctness condition and the security condition. These are formally described as follows.

$\textit{Correctness condition:}$ Each user needs to be able to fully recover the requested file from the signal transmitted by the server and its own cache content, i.e., 
\begin{align}\label{decode}
H\left(W_{d_k}\mid Z_k, X_{(d_{1},\ldots, d_{K})}\right)=0, \quad k \in [K].
\end{align}

$\textit{Security condition:}$ Each user cannot get any information about the files other than the one requested, i.e.,
\begin{align}\label{secureness}
I\left(\mathbf{W}_{[N] /\left\{d_{k}\right\}} ; Z_{k},X_{(d_{1},\ldots, d_{K})} \right)=0, \quad k \in [K].
\end{align}
The memory-rate pair $(M,R)$ is said to be \textit{securely achievable}, if for large
enough file size $F$, there exists a secure coded caching scheme. The optimal delivery rate $R^{\star}_S(M)$ given cache size $M$ is defined as
\begin{align}
R_{S}^{\star}(M)\mathrel{\triangleq}\inf\big\{R: (M,R) \text{ is securely achievable}\big\},\nonumber
\end{align}
where the infimum is over all secure coded caching schemes. 
We aim to characterize $R_{S}^{\star}(M)$, which is termed the optimal memory-rate tradeoff.

\section{Main Result}\label{section3}

The main results of this paper can be divided into two parts, the first part is characterizing the two end-points on the optimal memory-rate tradeoff $R_{S}^{\star}(M)$ for arbitrary number of users and files. More specifically, it has been shown that all securely achievable $(M,R)$ must satisfy $M \geq 1$ and $R \geq 1$ \cite{PCC2018}. Hence, we find the smallest worst-case delivery rate for the case of $M=1$, and the smallest cache size needed to achieve the worst-case delivery rate of $R=1$. And the second part of the main results is proposing a general achievable scheme, whose performance exceeds the best known existing schemes \cite{PCC2018}. Furthermore, the proposed achievable scheme is shown to be optimal when cache size is small in the two file case.  The main results are summarized in the following four theorems.

{\it Theorem 1 (end-point of the optimal memory-rate tradeoff curve at $M=1$):}
 	For $\forall N,K\ge 2$, in the case of $M=1$, the minimum worst-case delivery rate is
 	\begin{equation*}
 		R=\begin{cases}
 			K-1,&  \text{ if }\ N=2, \\
 			K,& \text{ if }\ N\ge 3.
 		\end{cases}
 	\end{equation*}
	In other words, we have $R_{S}^{\star}(1)=K-1$, when $N=2$, and $R_{S}^{\star}(1)=K$, when $N \geq 3$.

\begin{IEEEproof}The achievability of the case $N\ge 3$ follows from existing result \cite{PCC2018}. We show that when $N=2$, the existing achievability of $R=K$ is suboptimal\cite{PCC2018}, and we reduce the delivery rate by 1 through designing one user's cache content as a function of files and all other users' cache contents. In terms of the converse, we first derive two properties that any secure coded caching scheme must satisfy when $M=1$. The first property is that the cached content at a user is a deterministic function of the delivery signal and its requested message. This property further leads to the mutual independence property of a single file, and the cached content at each user. Then, the converse result follows from these two properties. The details of the proof can be found in Sections  \ref{Nan0127a} and \ref{Nan0127c}.
\end{IEEEproof}

\begin{Remark}
When cache size $M=1$, the result of Theorem 1 contains the optimal memory-rate tradeoff for $N=K=2$ \cite{PCC2018} as a special case,  which consists of only one corner point $(M,R)=(1,1)$.  For $K \geq 3$ and arbitrary $N$, existing achievability result states that rate $R = K$ is achievable, and existing converse result states that $R \geq \min\{\frac{N}{2}, K \}$ must be satisfied \cite{PCC2018}. From Theorem 1, we see that existing achievability result is tight for the case of $N \geq 3$ and arbitrary $K$, and existing converse result is tight for the case of $N \geq 2K$. In all other cases, the existing converse and achievability results are loose while Theorem 1 characterizes the optimal worst-case delivery rate. 
\end{Remark}
\begin{Remark}
	In the work\cite{FDsecure2022}, both file security and user privacy are considered, and it is shown that when \( N \geq 2K \), \( (M, R)=(1,K)\) is optimal. From our results, it can be concluded that even when only file security is considered, \( (M, R)=(1,K) \) is optimal under the condition \( N \geq 2K \).
\end{Remark}

{\it Theorem 2 (end-point of the optimal memory-rate tradeoff curve at $R=1$):}
 	For $\forall N,K\ge 2$, in the case of $R=1$, the minimum cache size achievable is
 	\begin{align}
 		M=(N-1)(K-1).\nonumber
 	\end{align}
	In other words, we have $R_{S}^{\star}((N-1)(K-1))=1$.

\begin{IEEEproof}	
The achievability mainly ensures that for any set of requests $(d_1, \cdots, d_K)$, each user $k$ has a combination of the same file $W$, same random key $\tilde{S}$ and the requested file $W_{d_k}$ in the cache content, i.e., $W_{d_k}\oplus W\oplus \tilde{S}$, where the size of both $W$ and $\tilde{S}$ is equal to the size of $W_{d_k}$. This way, when the server delivers the combination of the file $W$ and random key $\tilde{S}$, i.e., $ W\oplus \tilde{S}$, each user $k$ can use the cache content and delivery signal to recover its requested file $W_{d_k}$. In terms of the converse, we first derive two properties that any secure coded caching scheme must satisfy  when $R=1$. The first property is that the delivery signal is a deterministic function of the cached content of a user and its requested message. This property further leads to the mutual independence property of a single file, and some of the delivery signals. Then, the converse result follows from these two properties. The details of the proof can be found in Sections \ref{Nan0127b} and \ref{Nan0127d}.
\end{IEEEproof}
\begin{Remark}
For the case of delivery rate $R=1$, $K \geq 3$ and arbitrary $N$, existing achievability result states that memory size $M=N(K-1)$ is achievable, and existing converse result states that $M\ge{\lfloor{N/2}\rfloor}$ must be satisfied \cite{PCC2018}. From Theorem 2, we see that neither existing achievability nor converse result is tight, and Theorem 2 characterizes the minimum cache size achievable. 
\end{Remark}

{\it Theorem 3 (general achievable scheme):}
For $N\ge2,K\ge3,$  and cache size $M(t)=\frac{N t}{K-t}+1-\frac{1}{\binom{K}{t}-\binom{K-1}{t-1} }$  with  $t \in\{1, \ldots, K-2\}$ , the following rate is secretively achievable
\begin{align}
	R(M(t)) = \frac{K}{t+1}.
\end{align}

\begin{IEEEproof}	
	The proposed achievability scheme is based on the observation that the scheme in \cite{PCC2018} utilizes both secret sharing and a set of independently generated random keys. However,  the implementation of secret sharing itself also generates random keys, leading to a redundancy of random key. Our scheme achieves better performance by reducing this random key redundancy. The details of the proof can be found in Section \ref{section43} .
\end{IEEEproof}

\begin{Remark}
	Compared to existing schemes, i.e., those in \cite{PCC2018}, the scheme proposed in Theorem 3 requires a smaller cache size to achieve the same rate. More specifically, to achieve the same rate, the cache size required by our scheme is \(\frac{1}{\binom{K}{t} - \binom{K-1}{t-1}}\) smaller.
\end{Remark}
\begin{Remark}
	The convex envelope of the achievable $(M,R)$ points given by Theorems 1, 2, and 3 is achievable. Note that in Theorem 3, when \( N = 2 \) and $K\ge 4$, an achievable scheme can be obtained by setting \( t = K - 2 \). However, it is not as good as the the cache-sharing scheme between the scheme corresponding to \( t = K - 3 \) in Theorem 3 and the achievable scheme in Theorem 2.
\end{Remark}

{\it Theorem 4 (optimal memory-rate tradeoff curve for small cache size, $N=2$ and arbitrary $K$):}
	For \(N = 2\) and \( K \geq 3\), when \(M \in \left[1, \frac{K}{K-1}\right]\), the optimal memory-rate tradeoff is given by
	\begin{align}
		(K-1)(K-2)M+2R\ge K(K-1).\label{theoeq4}
	\end{align}
	 In other words, $R_{S}^{\star}(M)=\frac{1}{2} \left(K(K-1)-(K-1)(K-2)M \right)$, for $M \in \left[1, \frac{K}{K-1}\right]$, $N=2$, and $K \geq 3$.

\begin{IEEEproof}	
	The achievability scheme can be derived from Theorem 3. The main idea of the converse proof is derived by utilized the symmetry property in the coded caching problem \cite{tianchao}. The details of the converse proof can be found in Section \ref{section53}.
\end{IEEEproof}

\begin{figure*}[t]
	\centering
	\begin{subfigure}[b]{0.45\textwidth}
		 \centering
		\includegraphics[width=\textwidth]{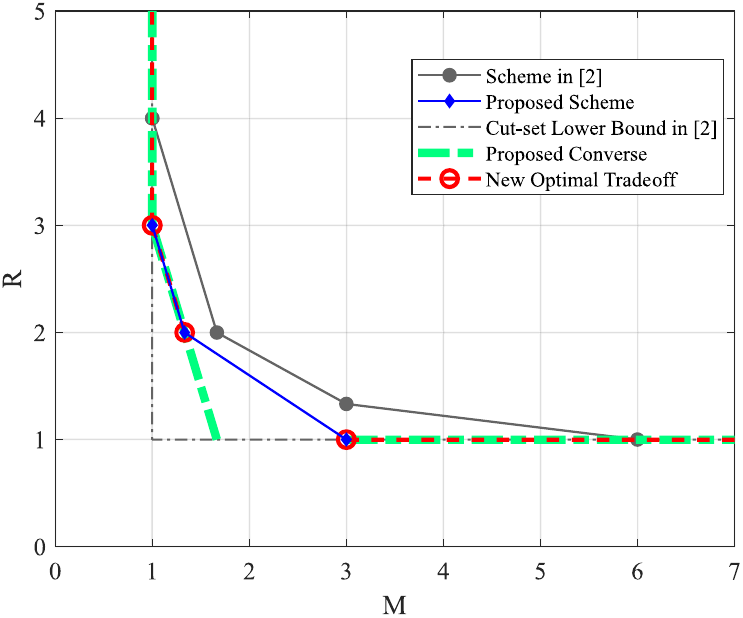}
		\caption{$N=2,K=4$}
		\end{subfigure}
		 \hfill
		 \begin{subfigure}[b]{0.45\textwidth}
		 \centering
		 \includegraphics[width=\textwidth]{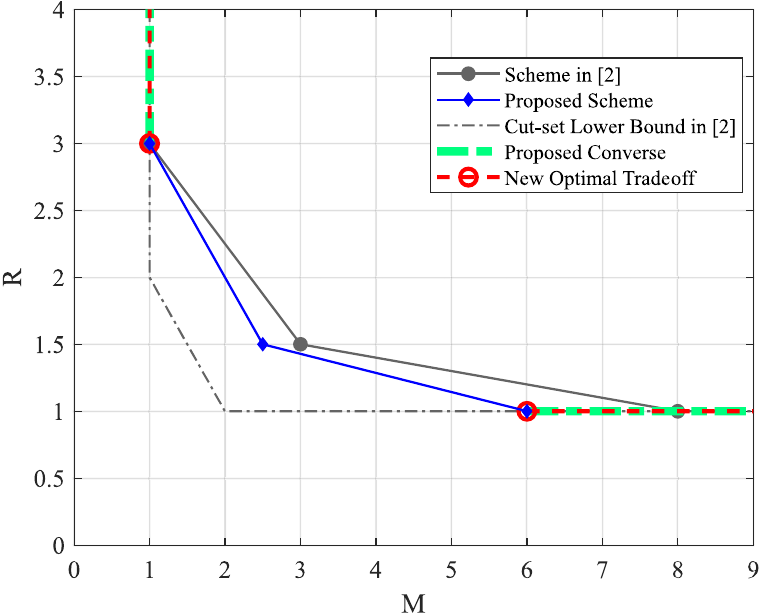}
		 \caption{$N=4,K=3$}
		 \end{subfigure}
	\caption{Comparison of existing results\cite{PCC2018} and results of this paper.}
	\label{fig: figure}
\end{figure*}

\begin{Remark}
 To demonstrate the improvement of our main results over existing results, we provide a couple of examples in Fig. \ref{fig: figure}. The horizontal axis represents the cache size \(M\) of each user, while the vertical axis represents the delivery rate $R$ in the worst-case scenario. The memory-rate tradeoff achieved by existing achievable schemes\cite{PCC2018} is depicted by the  black solid line with circular markers, while the existing converse result\cite{PCC2018} is represented by the black dash-dotted line. Our newly proposed scheme is shown with the blue solid line with diamond markers, and our newly proposed converse result is shown with the thick dash-dotted green line. The red dashed line with hollow circular markers depicts the optimal memory-rate tradeoff, i.e., where our proposed achievability and converse results meet. For the case where the number of files \(N = 2\) and the number of users \(K = 4\), the new schemes proposed in Theorem 1, i.e., $(M,R)=(1,3)$, Theorem 2, i.e., $(M,R)=(3,1)$,  and Theorem 3, i.e., $(M,R)=(4/3,2)$ outperform the existing schemes\cite{PCC2018}, i.e., $(M,R)=(1,4),(5/3,2),(3,4/3),(6,1)$. The new converse results derived from Theorem 1, i.e., $R\ge 2$ for $M=1$, Theorem 2, i.e., $M\ge3$ for $R=1$, and Theorem 4, i.e., $3M+R\ge 6$ provide an improvement over the existing converse\cite{PCC2018}, i.e., $M\ge 1,R\ge1$. The new optimal tradeoff results here are $R \geq 3$ for $M=1$, $3M+R\ge 6$ for $M\in[1,4/3]$ and $M\ge3$ for $R=1$.   For the case where \(N = 4\) and \(K = 3\), the new schemes proposed in Theorem 2, i.e., $(M,R)=(5/2,3/2)$ and Theorem 3, i.e., $(M,R)=(6,1)$ outperform the existing schemes\cite{PCC2018}, i.e., $(M,R)=(3,3/2),(8,1)$. The new converse results derived from Theorem 1, i.e., $R\ge 3$ for $M=1$ and Theorem 2, i.e., $M\ge 6$ for $R=1$, provide an improvement over the existing converse\cite{PCC2018} in the case of $M=1$ and $R=1$, i.e., $R\ge2$ for $M=1$ and $M\ge2$ for $R=1$. 
\end{Remark}

\section{Achievability Proofs}\label{section4}
In this section, we will provide the achievability proofs of Theorems 1 and 2 and the proof of Theorems 3. More specifically, we will provide three achievable schemes, one works for the case of small buffer, i.e., $M=1$, one works for the case of small rate, i.e., $R=1$, and the other works for the case of general cache size, i.e., $M>1,R>1$. 
 
Note that when all users request the same file, sending the file requested by the users directly satisfies the correctness and security conditions. This incurs a delivery rate of $1$ and is \emph{not} the worst-case delivery rate\cite{PCC2018}. Therefore, in the following description of achievable schemes, without loss of generality, we only consider the case where the users do not all request the same file. 

\subsection{Achievability proof of Theorem 1}
\label{Nan0127a}
When $N\ge 3$, adopting the achievable scheme proposed in \cite{PCC2018}, the worst-case delivery rate of $K$ is achievable. For completeness, we briefly write the scheme here. The randomness generated at the server is $S=(S_1, \cdots, S_K)$, where the $S_k$s are i.i.d., and each consists of $F$ i.i.d. symbols uniformly distributed on $\mathbb{F}_{q}$, which means that the amount of randomness $L=K$. The server places $S_k$ in the $k$-th user's cache during the placement phase, i.e., $Z_k=S_k$. During the delivery phase, the server delivers $W_{d_k} \oplus S_k$, for $k \in [K]$. It can be easily seen that the correctness and security conditions are satisfied, and the worst-case delivery rate is $R=K$. 
 
Hence, we only need to prove the case of two files, i.e., $N=2$. Before presenting the proposed general achievable scheme for $N=2$ and arbitrary $K$, we first start with a motivating example. 
 
\subsubsection{Motivating example $N=2, K=3$ and $M=1$}\hfill \break
\indent The finite field is taken to be $\mathbb{F}_{3}$, i.e., $q=3$. During the placement phase, the following are cached in each user's cache respectively, 
\begin{align*}
	Z_1&=\{S_1\},\\\ Z_2&=\{S_2\},\\\ Z_3&=\{2W_1\oplus2W_2\oplus S_1\oplus S_2\}, 
\end{align*}
where the randomness generated at the server, i.e., $S$, is split into 2 parts, each consists of $F$ symbols, i.e., $S=(S_1,S_2)$ and $L=2$.

In order to design the delivery signal, it is necessary to find a triple of coefficients $(a_1,a_2,a_3) \in \mathbb{F}_{3}$ with non-zero elements for every demand $(d_1,d_2,d_3)$ satisfying 
\begin{align}
	{W_1} \oplus {W_2}={a _1} {W_{d_1}}\oplus {a _2}{W_{d_2}}\oplus {a _3}{W_{d_3}},\label{coecons}
\end{align}
under $\mathbb{F}_{3}$. For example, when $(d_1,d_2,d_3)=(1,1,2)$, take $(a_1,a_2,a_3)=(2,2,1)$, when $(d_1,d_2,d_3)=(2,1,2)$, take $(a_1,a_2,a_3)=(2,1,2)$, and we can find the corresponding $(a_1,a_2,a_3)$ for other $(d_1,d_2,d_3)$s by symmetry.

During the delivery phase, the server delivers
 \begin{align}
 X_{(d_1,d_2,d_3)}=	\{a_1W_{d_1}\oplus 2S_1, a_2W_{d_2}\oplus 2S_2 \}.\nonumber
 \end{align}
 
We first show that the correctness condition (\ref{decode}) is satisfied. From $a_1W_{d_1}\oplus 2S_1$ received from the server and the cache content $S_1$, the first user possesses two linear independent combinations of $W_{d_1}$ and $S_1$, and therefore can decode both $W_{d_1}$ and $S_1$. Similarly, by symmetry, user 2 can decode both $W_{d_2}$ and $S_2$. 
As for the third user, its decoding function is
\begin{subequations}
\begin{align}
 	&(2W_1\oplus2W_2\oplus S_1\oplus S_2)\oplus (a_1W_{d_1}\oplus 2S_1)\oplus (a_2W_{d_2}\oplus 2S_2)\notag\\
 	&=(2W_1\oplus2W_2)\oplus (a_1W_{d_1}\oplus a_2W_{d_2})\notag\\
 	&=(2W_1\oplus2W_2)\oplus (W_1\oplus W_2)\oplus(2a_3W_{d_3})\label{6a}\\
 	&=2a_3W_{d_3}\notag,
\end{align}
\end{subequations}
where (\ref{6a}) follows from (\ref{coecons}).  Hence, the correctness condition (\ref{decode}) is satisfied for all three users. 
 
For the security condition (\ref{secureness}), we have
\begin{align}
	I\left(W_{[N] /\left\{d_{k}\right\}} ;  Z_{k},X_{(d_1,d_2,d_3)} \right)
	&=H\left(Z_{k},X_{(d_1,d_2,d_3)}\right)-H\left(Z_{k},X_{(d_1,d_2,d_3)} |W_{[N] /\{d_{k}\}}\right)\notag\\
	&=(3-3)F=0, \quad \forall k \in [K].\nonumber
\end{align}	
Hence, the proposed scheme satisfies both the correctness condition (\ref{decode}) and the security condition (\ref{secureness}), and it has the worst-case delivery rate of $R=2$. 
 
 \subsubsection{General achievable scheme for $M=1$, $N=2$ and arbitrary $K$}\label{section411}\hfill \break
\indent The finite field is taken to be $\mathbb{F}_{3}$, i.e., $q=3$. Split the randomness $S$ generated at the server into $K-1$ parts, each consists of $F$ symbols, i.e., $S=(S_1,S_2, \cdots, S_{K-1})$, and the amount of randomness $L=K-1$.
Let
\begin{align}
	\mathbf{v}=\begin{bmatrix}W_1\ W_2\ S_1 \ \cdots \ S_{K-1}\end{bmatrix}^T.\nonumber
\end{align}
Then the cached contents at the users are
\begin{align*}
  	Z_k&=\begin{bmatrix}0\ 0  \ \mathbf{e}^T_k \end{bmatrix}\mathbf{v} \triangleq  \mathbf{G}_{Z_k}\mathbf{v},\qquad k  \in [K-1],\\
	Z_K&=\begin{bmatrix}2\ 2  \ \bigoplus_{i=1}^{K-1}\mathbf{e}^T_i\end{bmatrix}\mathbf{v} \triangleq \mathbf{G}_{Z_K}\mathbf{v}.
\end{align*}
For a certain demand $(d_1,\ldots,d_k)$ where not all users request the same file, 
\begin{align*}
	X_{(d_1, \cdots, d_K)}=\begin{bmatrix}\mathbf{A}_{(K-1)\times 2}\ 2\mathbf{I}_{(K-1)\times (K-1)}\end{bmatrix}\mathbf{v} \triangleq \mathbf{G}_X\mathbf{v},
\end{align*}
where
$\mathbf{A}$ corresponds to the coefficient matrix of the file in the delivery signal. We take $\mathbf{A}$ to be of the form
\begin{align*}
\mathbf{A}=\left[
\begin{array}{cc}
\delta_{1d_1}a_1 & \delta_{2d_1}a_1 \\
\vdots & \vdots\\
\delta_{1d_{K-1}}a_{K-1} & \delta_{2d_{K-1}}a_{K-1}\\
\end{array}\right],
\end{align*}
where \begin{equation}
	\delta_{ij} \triangleq \begin{cases}
		1& \text{ if } \ i=j, \\
		0& \text{ if }\ i\ne j.
	\end{cases} \label{Nan0127f}
\end{equation}
In other words, if the $k$-th user request $W_1$, i.e., $d_k=1$, then $(\delta_{1d_k},\delta_{2d_k})=(1,0)$, and if the $k$-th user request $W_2$, i.e.. $d_k=2$, then $(\delta_{1d_k},\delta_{2d_k})=(0,1)$, $k \in [K-1]$. Let the set of users requesting $W_1$ be denoted as $\mathcal{K}_1=\{k|d_k=1,k\in [K]\} \triangleq \{l_1, l_2, \cdots, l_{|\mathcal{K}_1|}\}$, then, $(a_{l_1}, a_{l_2}, \cdots, a_{l_{|\mathcal{K}_1|}})$ take values as follows: the first pair takes the value of $(1,2)$, the second pair again takes the value of $(1,2)$, and so on and so forth, until there is only one or two elements left. When there is only one element left, it takes the value of $1$ and when there are two elements left, they take the value of $(2,2)$. The same operation is done for $a_{k}$, $k \in \mathcal{K}_2=\{k|d_k=2,k\in [K]\}$. 
Thus, we have that $a_i \neq 0$ for all $i \in [K]$, and
\begin{align}
\textstyle\bigoplus_{i=1}^{K}\delta_{1d_i}a_{i}=\textstyle\bigoplus_{i=1}^{K}\delta_{2d_i}a_{i}=1\label{matrixconstrain}. 
\end{align}
Now that we have described the caching function and encoding function of the proposed scheme, we will prove that the correctness condition (\ref{decode})  and the security condition (\ref{secureness}) are satisfied in Appendix \ref{app2}.

 \subsection{Achievability proof of Theorem 2} 
 \label{Nan0127b}
We propose a new achievable scheme for the case of $R=1$, and arbitrary $N$ and $K$. We again start with a motivating example.

\subsubsection{Motivating example $N=3,K=3$ and $R=1$}\hfill \break
\indent The finite field is taken to be $\mathbb{F}_{2}$, i.e., $q=2$. Let the randomness generated at the server, i.e., $S$, be split into 4 parts, each consists of $F$ symbols, i.e., $S=(S_{1,1},S_{1,2}, S_{2,1}, S_{2,2})$, and $L=4$. During the placement phase, the following are cached in each user's cache, respectively, 
\begin{align*}
	&Z_1=\left\{ 
	\begin{matrix}
		W_1\oplus W_2 \oplus S_{1,1}, S_{1,2}, W_1\oplus W_3 \oplus S_{2,1}, S_{2,2} 
	\end{matrix}\right\},\\
	&Z_2=\left\{ 
	\begin{matrix}
		S_{1,1}, W_1\oplus W_2 \oplus S_{1,2}, S_{2,1}, W_1\oplus W_3 \oplus S_{2,2} 
	\end{matrix}\right\},\\
	&Z_3=\left\{ 
	\begin{matrix}
		S_{1,1},  S_{1,2},  S_{2,1},  S_{2,2} \\
	\end{matrix}\right\}.
\end{align*}
Note that the cache size is $(N-1)(K-1)=4$. 

During the delivery phase, for every demand $(d_1,d_2,d_3)$, the server delivers
\begin{align*}
	X_{(d_1,d_2,d_3)}=	W_{d_3}\textstyle\bigoplus_{i=1}^{2} (S_{d_3-1,i}\oplus S_{d_{i}-1,i}),
\end{align*}
where we define $S_{0,i}=0$.

We first show that the correctness condition (\ref{decode}) is satisfied. The correctness condition (\ref{decode}) is obvious for the third user, as the random keys $(S_{1,1},S_{1,2}, S_{2,1}, S_{2,2})$ in the delivery signal are all cached by the third user. 
For the first two users, note that for $\forall n\in\{1,2,3\},k,k'\in \{1,2\},k\ne k'$, according to placement, we have $W_1\oplus W_{n}\oplus S_{n-1,k}\in Z_k$ and $S_{n-1,k'}\in  Z_k$. Hence, 
for any demand $(d_1,d_2,d_3)$ and $k,k'\in \{1,2\},k\ne k',$
$\{W_1\oplus W_{d_3}\oplus S_{d_3-1,k},W_1\oplus W_{d_i}\oplus S_{d_i-1,k}\}\in Z_k$, and $ \{S_{d_3-1,k'},S_{d_1-1,k'}\}\in Z_k$, 
which leads to
\begin{align}
	W_{d_3}\oplus W_{d_k}\oplus S_{d_3-1,k}\oplus S_{d_k-1,k}&\in Z_k, \label{wwss}\\
	S_{d_3-1,k'}\oplus S_{d_i-1,k'}&\in Z_k.\label{ss}
\end{align}
 Then, for the first two users, from (\ref{wwss}) and (\ref{ss}), we have
\begin{align*}
	W_{d_3}\oplus W_{d_k} \textstyle\bigoplus_{i=1}^{2} (S_{d_3-1,i}\oplus S_{d_{i}-1,i})\in Z_k,
\end{align*}
which is equal to $X_{(d_1,d_2,d_3)}\oplus W_{d_k}\in Z_k$ for $k\in\{1,2\}$. Therefore, the first two users can decode their requested file from the local cache content and the delivery signal. Hence, the correctness condition (\ref{decode}) is satisfied.

For the security condition (\ref{secureness}), we have
\begin{align*}
	I\left(W_{[N] /\left\{d_{k}\right\}} ;  Z_{k},X_{(d_1,d_2,d_3)} \right)
	&=H\left(Z_{k},X_{(d_1,d_2,d_3)}\right)-H\left(Z_{k},X_{(d_1,d_2,d_3)} |W_{[N] /\{d_{k}\}}\right)\notag\\
	&=(5-5)F=0, \quad \forall k=1,\cdots, K.
\end{align*}
Hence, the proposed scheme satisfies both the correctness condition (\ref{decode}) and the security condition (\ref{secureness}), and it has the worst-case delivery rate of $R=1$ and cache size $M=(N-1)(K-1)=4$.

\subsubsection{General Achievable Scheme for $R=1$ and arbitrary $N$ and $K$} \hfill \break
\indent The finite field is taken to be $\mathbb{F}_{2}$, i.e., $q=2$. Split the randomness $S$ generated at the server into $(N-1)(K-1)$ parts, each consists of $F$ symbols, i.e., $S=(S_{1,1},\cdots, S_{N-1,K-1})$ and $L=(N-1)(K-1)$. 
Let
\begin{align*}
	\mathbf{v}=\begin{bmatrix}W_1\ \cdots\ W_N\ S_{1,1} \ \cdots\ S_{N-1,K-1}\end{bmatrix}^T.
\end{align*}
During the placement phase, for $\forall n\in [N],k,k'\in [K-1],k\ne k'$,  user $k$ caches $W_1\oplus W_{n}\oplus S_{n-1,k}$ and $S_{n-1,k'}$, where we define $S_{0,i}=0$, which does not need to be cached. User $K$ caches all of the $(N-1)(K-1)$ random keys, i.e., $\{S_{1,1},\cdots, S_{N-1,K-1}\}$.  
	
Written in matrix form, we have 	
	for $k \in [K-1]$, 
	\begin{align}
		Z_k=\left[
		\begin{array}{cc}
			\mathbf{e}'_{k}(\mathbf{e}_{1}^T\oplus\mathbf{e}_{2}^T)&\textstyle\bigoplus_{j\in[K-1]}\mathbf{e}'_{j}\mathbf{e}^T_{1j}\\
			\mathbf{e}'_{k}(\mathbf{e}_{1}^T\oplus\mathbf{e}_{3}^T)& \textstyle\bigoplus_{j\in[K-1]}\mathbf{e}'_{j}\mathbf{e}^T_{2j}\\
			\multicolumn{2}{c}{\cdots} \\
			\mathbf{e}'_{k}(\mathbf{e}_{1}^T\oplus\mathbf{e}_{N}^T)& \textstyle\bigoplus_{j\in[K-1]}\mathbf{e}'_{j}\mathbf{e}^T_{N-1j}
		\end{array}\right]\mathbf{v}\triangleq\mathbf{G}_{Z_k}\mathbf{v},\nonumber
	\end{align}
and for the $K$-th user, its cached content is
	\begin{align}
		Z_K=\left[
		\begin{array}{cc}
			\mathbf{0}'&\textstyle\bigoplus_{j\in[K-1]}\mathbf{e}'_{j}\mathbf{e}^T_{1j}\\
			\mathbf{0}'& \textstyle\bigoplus_{j\in[K-1]}\mathbf{e}'_{j}\mathbf{e}^T_{2j}\\
			\multicolumn{2}{c}{\cdots} \\
			\mathbf{0}'& \textstyle\bigoplus_{j\in[K-1]}\mathbf{e}'_{j}\mathbf{e}^T_{N-1j}
		\end{array}\right]\mathbf{v}=\begin{bmatrix}\mathbf{0}\ \mathbf{I}\end{bmatrix}\mathbf{v}\triangleq\mathbf{G}_{Z_K}\mathbf{v},\nonumber
	\end{align}
where $\mathbf{e}'_i$ denotes a column vector with dimension $K-1$ whose $i$-th element is $1$ and the other elements are 0, $\mathbf{e}_i$ is of dimension $N\times 1$,  $\mathbf{e}_{ij}$ represents a column vector of dimension $(N-1)(K-1)$, where only the $((i-1)(K-1)+j)$-th element is 1 and the other elements are 0. Hence,  $\mathbf{e}'_{j}\mathbf{e}^T_{1j}$ represents a matrix of dimension $(K-1)\times [(N-1)(K-1)]$, where the $(j, (i-1)(K-1)+j)$-th element is 1, and the other elements are 0. $\mathbf{0}',\mathbf{0}$ represents the all-zero matrix with dimension $(K-1)\times N$ and $[(N-1)(K-1)]\times N$,  respectively,  and $\mathbf{I}$ is the identity matrix of size $(N-1)(K-1)$.

During the delivery phase, for a demand $(d_1,\ldots,d_k)$ where not all users request the same file, server delivers 	$X_{(d_1,\ldots,d_k)}=	W_{d_K}\textstyle\bigoplus_{i=1}^{K-1} (S_{d_K-1,i}\oplus S_{d_{i}-1,i})$, which is equal to
\begin{align}
		X_{(d_1,\ldots,d_k)}=\begin{bmatrix}\mathbf{e}_{d_K}^T\ \textstyle\bigoplus_{i=1}^{K}(\mathbf{e}^T_{d_K-1,i}\oplus \mathbf{e}^T_{d_i-1,i})\end{bmatrix}\mathbf{v}\triangleq\mathbf{G}_X\mathbf{v},\nonumber
\end{align}
where we define $\mathbf{e}_0^T=\mathbf{0}^T$ with dimension $1\times N$, and $\mathbf{e}_{i,K}^T=\mathbf{0}^T$ with dimension $1 \times [(N-1)(K-1)]$.

Now that we have described the caching function and encoding function of the proposed scheme, we show that the correctness condition (\ref{decode}) and the security condition (\ref{secureness}) are satisfied in Appendix \ref{app3}.

\subsection{Achievability proof of Theorem 3, i.e., for $M>1$  and  $R>1$ } \label{section43}
We want to prove thar for the cache size $M(t)=\frac{N t}{K-t}+1-\frac{1}{\binom{K}{t}-\binom{K-1}{t-1} }$  with  $t \in\{1, \ldots, K-2\}$, the following rate is securely achievable,
\begin{align}
	R(M(t)) = \frac{K}{t+1}.
\end{align} 

Our proposed scheme is built on shares which satisfy the property of non-standard secret sharing, which is defined as follows. 

{\it Definition 1:} For \( m < n \), we say that the \( n \) equal-sized shares \( T_1, \cdots, T_n \) satisfy the \((m,n)\) \emph{non-standard secret sharing} property for a uniformly distributed secret \( W \)\cite{PCC2018}, if 
any \( m \) shares do not reveal any information about the secret $W$, and all \( n \) shares together can fully reveal the secret $W$, i.e.,
\begin{align*}
I\left(W ; T_{\mathcal{M}}\right) & =0, \quad\forall \mathcal{M} \subset[n], |\mathcal{M}|=m, \\
H\left(W|T_{[n]}\right) & =0 .
\end{align*}
It has been shown in \cite{PCC2018} that the shares must have a size of at least \(\frac{\log|\mathcal{W}|}{n-m}\) bits, where $\mathcal{W}$ is the alphabet of the secret $W$.

Before introducing the general scheme, we first start with a motivating example.

\subsubsection{Motivating example $N=3,K=3$ and $M=2$, i.e., $t=1$}\hfill \break
\indent The finite field is taken to be $\mathbb{F}_{3}$, i.e., $q$ is taken to be $3$. For $K=3$ and $t=1$, we first propose a scheme which generates shares that satisfy the $(1, 3)$ non-standard secret sharing property. We will then utilize these shares to build the proposed achievable scheme. More specifically,  for all $k \in [K]$, 
the file $W_k$ is divided into two equal parts, denoted as \(W_k^1\) and \(W_k^2\), each of size \(\frac{F}{2}\). A random key \(S_{k}\) of size \(\frac{F}{2}\) is generated, and the three shares $T_k^{\{1\}}, T_k^{\{2\}}, T_k^{\{3\}}$ are formed as follows 
\begin{align}
 T_k^{\{1\}} = W_k^1 \oplus_3 S_{k}, \quad \ T_k^{\{2\}} = W_k^2 \oplus_3 S_{k},\quad \ T_k^{\{3\}} = S_{k}. \label{Nan241224a}
 \end{align} 
 It can be checked that the three shares satisfy the $(1,3)$ non-standard secret sharing property for $W_k$, i.e., any 1 share will not reveal any information about the secret $W_k$, but all 3 shares will fully reveal $W_k$. In addition, the server needs to generate three additional random keys, each of size \(\frac{F}{2}\), denoted as $\{S_\mathcal{V},\mathcal{V}\subset[3],|\mathcal{V}|=2\}$, i.e., $\{S_{\{1,2\}},S_{\{1,3\}},S_{\{2,3\}}\}$. Therefore, the total size of the randomness $LF$ generated by the server is $\frac{F}{2} + \frac{3F}{2} = 2F$.	
	
The proposed achievable scheme is built on the above shares \(\{T_k^{\{i\}}\}_{k \in [K], i \in [3]}\), and the fact that for a given $k \in [K]$, the three shares \(\{T_k^{\{i\}}\}_{i \in [3]}\) satisfy the $(1,3)$ non-standard secret sharing property for $W_k$, will be used in the proof of the  (\ref{decode}) and the security condition (\ref{secureness}). 
More specifically, during the placement phase, the cached contents are
\begin{align*}
	Z_1=&\{T_1^{\{1\}},T_2^{\{1\}},T_3^{\{1\}},S_{\{1,3\}}\},
	\\Z_2=&\{T_1^{\{2\}},T_2^{\{2\}},T_3^{\{2\}},S_{\{2,3\}}\},
	\\Z_3=&\{T_1^{\{3\}}\oplus_3 S_{\{1,3\}},T_2^{\{3\}}\oplus_3 S_{\{1,3\}}, T_3^{\{3\}}\oplus_3 S_{\{1,3\}}, S_{\{2,3\}}\ominus_3 S_{\{1,3\}} \}.
\end{align*}

During the delivery phase, for a demand $(d_1,d_2,d_3)$ where not all users request the same file, the server delivers 
\begin{align*}
	X_{(d_1,d_2,d_3)}&=\{2S_{\{1,3\}}\oplus_3 T_{d_1}^{\{3\}}\oplus_3 T_{d_3}^{\{1\}}, 2S_{\{2,3\}}\oplus_3 T_{d_2}^{\{3\}}\oplus_3 T_{d_3}^{\{2\}},T_{d_1}^{\{2\}}\oplus_3 T_{d_2}^{\{1\}}\}.
\end{align*}

The correctness condition (\ref{decode}) can be satisfied for the following reasons. The first user with cached content $Z_1$ can use $T_{d_2}^{\{1\}}$ to decode $T_{d_1}^{\{2\}}$ from $T_{d_1}^{\{2\}}\oplus_3 T_{d_2}^{\{1\}}\in X_{(d_1,d_2,d_3)}$, and use $T_{d_3}^{\{1\}},S_{\{1,3\}}$ to decode $T_{d_1}^{\{3\}}$ from $S_{\{1,3\}}\oplus_3 T_{d_1}^{\{3\}}\oplus_3 T_{d_3}^{\{1\}}\in X_{(d_1,d_2,d_3)}$. Then, the first user has the knowledge of all \( 3 \) shares $\{T_{d_1}^{\{1\}},T_{d_1}^{\{2\}},T_{d_1}^{\{3\}}\}$, which together can fully reveal $W_{d_1}$. The second user with cached content $Z_2$ can reveal $W_{d_2}$ similar to the first user with cached content $Z_1$. The third user with cached content $Z_3$ can use $T_{d_1}^{\{3\}}\oplus_3 S_{\{1,3\}}$ to decode $T_{d_3}^{\{1\}}\oplus_3 S_{\{1,3\}}$ from $2S_{\{1,3\}}\oplus_3 T_{d_1}^{\{3\}}\oplus_3 T_{d_3}^{\{1\}}\in X_{(d_1,d_2,d_3)}$, and use $T_{d_2}^{\{3\}}\oplus_3 S_{\{1,3\}},S_{\{2,3\}}\ominus_3 S_{\{1,3\}}$ to decode $T_{d_3}^{\{2\}}\oplus_3 S_{\{1,3\}}$ from $2S_{\{2,3\}}\oplus_3 T_{d_2}^{\{3\}}\oplus_3 T_{d_3}^{\{2\}}\in X_{(d_1,d_2,d_3)}$. At this point, the third user with cached content $Z_3$ can obtain three shares  each combined with a random key modulo operation, i.e., $\{T_{d_3}^{\{1\}}\oplus_3 S_{\{1,3\}},T_{d_3}^{\{2\}}\oplus_3 S_{\{1,3\}},T_{d_3}^{\{3\}}\oplus_3 S_{\{1,3\}}\}$. Next, we prove that for the shares constructed in (\ref{Nan241224a}), the third user can decode \(W_{d_3}\). More specifically, from (\ref{Nan241224a}), the third user has the knowledge of $\{W_{d_3}^1 \oplus_3 S_{{d_3}}\oplus_3 S_{\{1,3\}},W_{d_3}^2 \oplus_3 S_{{d_3}}\oplus_3 S_{\{1,3\}},S_{{d_3}}\oplus_3 S_{\{1,3\}}\}$. Thus, \(W_{d_3}^1\) and \(W_{d_3}^2\) can be decoded using \(S_{{d_3}} \oplus_3 S_{\{1,3\}}\), thereby revealing \(W_{d_3}\). Hence, we have shown that all of the users can decode their requested files, i.e., the correctness condition (\ref{decode}) is satisfied.

The security condition (\ref{secureness}) can be satisfied for the following reasons. For the first user with cached content $Z_1$, we have
\begin{subequations}
	\begin{align}
		&I(W_{[3]/ d_1};Z_1,X_{(d_1,d_2,d_3)})\notag
		\\&=I(W_{[3]/ d_1};\{T_{d_1}^{\{1\}},T_{d_1}^{\{2\}},T_{d_1}^{\{3\}}\},Z_1,X_{(d_1,d_2,d_3)})\label{Nan241224b}
		\\&= I(W_{[3]/ d_1}; \{T_{d_1}^{\{1\}},T_{d_1}^{\{2\}},T_{d_1}^{\{3\}}\}\cup\{T_1^{\{1\}},T_2^{\{1\}},T_3^{\{1\}},S_{\{1,3\}}\}\notag
		\\&\qquad\cup\{2S_{\{1,3\}}\oplus_3 T_{d_1}^{\{3\}}\oplus_3 T_{d_3}^{\{1\}}, 2S_{\{2,3\}}\oplus_3 T_{d_2}^{\{3\}}\oplus_3 T_{d_3}^{\{2\}},T_{d_1}^{\{2\}}\oplus_3 T_{d_2}^{\{1\}}\})\notag
		\\&= I(W_{[3]/ d_1}; \{T_{d_1}^{\{1\}},T_{d_1}^{\{2\}},T_{d_1}^{\{3\}}\}\cup\{T_1^{\{1\}},T_2^{\{1\}},T_3^{\{1\}},S_{\{1,3\}}\}\cup\{ 2S_{\{2,3\}}\oplus_3 T_{d_2}^{\{3\}}\oplus_3 T_{d_3}^{\{2\}}\})\label{133b}
		\\&= I(W_{[3]/ d_1}; \{T_{d_1}^{\{1\}},T_{d_1}^{\{2\}},T_{d_1}^{\{3\}}\}\cup\{T_1^{\{1\}},T_2^{\{1\}},T_3^{\{1\}}\})\label{133c}
		\\&=0,\label{secure11a}
	\end{align}
\end{subequations}
where (\ref{Nan241224b}) follows from the fact that the first user can decode $\{T_{d_1}^{\{1\}},T_{d_1}^{\{2\}},T_{d_1}^{\{3\}}\}$, as discussed in the correctness condition, (\ref{133b}) follows because the terms $2S_{\{1,3\}} \oplus_3 T_{d_1}^{\{3\}} \oplus_3 T_{d_3}^{\{1\}}$ and $T_{d_1}^{\{2\}} \oplus_3 T_{d_2}^{\{1\}}$ are deterministic functions of  \(\{T_{d_1}^{\{1\}}, T_{d_1}^{\{2\}}, T_{d_1}^{\{3\}}\} \cup \{T_1^{\{1\}}, T_2^{\{1\}}, T_3^{\{1\}}, S_{\{1,3\}}\}\), 
 (\ref{133c}) holds because the random keys, $S_{\{1,3\}},S_{\{2,3\}},$ are generated independently of everything else, and (\ref{secure11a}) follows from the fact that the three shares \(\{T_k^{\{i\}}\}_{i \in [3]}\) satisfy the $(1,3)$ non-standard secret sharing property for $W_k$, and the fact that there is no other file with more than one share, aside from \( W_{d_1} \), in the mutual information. For the second user with cached content \(Z_2\), the security condition can be similarly derived. For the last user with cached content $Z_3$, we have
\begin{subequations}
	\begin{align}
		&I(W_{[3]/ d_3};Z_3,X_{(d_1,d_2,d_3)})\notag
		\\&=I(W_{[3]/ d_3};\{T_{d_3}^{\{1\}}\oplus_3 S_{\{1,3\}},T_{d_3}^{\{2\}}\oplus_3 S_{\{1,3\}}, T_{d_3}^{\{3\}}\oplus_3 S_{\{1,3\}} \},Z_3,X_{(d_1,d_2,d_3)})\label{14a}
		\\&=I(W_{[3]/ d_3};\{T_{d_3}^{\{1\}}\oplus_3 S_{\{1,3\}},T_{d_3}^{\{2\}}\oplus_3 S_{\{1,3\}}, T_{d_3}^{\{3\}}\oplus_3 S_{\{1,3\}} \}\notag
		\\&\qquad\cup\{T_{d_1}^{\{3\}}\oplus_3 S_{\{1,3\}},T_{d_2}^{\{3\}}\oplus_3 S_{\{1,3\}},S_{\{2,3\}}\ominus_3 S_{\{1,3\}} \}\notag
		\\&\qquad \cup\{2S_{\{1,3\}}\oplus_3 T_{d_1}^{\{3\}}\oplus_3 T_{d_3}^{\{1\}}, 2S_{\{2,3\}}\oplus_3 T_{d_2}^{\{3\}}\oplus_3 T_{d_3}^{\{2\}},T_{d_1}^{\{2\}}\oplus_3 T_{d_2}^{\{1\}}\})\notag
		\\&=I(W_{[3]/ d_3};\{T_{d_3}^{\{1\}}\oplus_3 S_{\{1,3\}},T_{d_3}^{\{2\}}\oplus_3 S_{\{1,3\}}, T_{d_3}^{\{3\}}\oplus_3 S_{\{1,3\}} \}\notag
		\\&\qquad\cup\{T_{d_1}^{\{3\}}\oplus_3 S_{\{1,3\}},T_{d_2}^{\{3\}}\oplus_3 S_{\{1,3\}}, S_{\{2,3\}}\ominus_3 S_{\{1,3\}} \}\cup\{T_{d_1}^{\{2\}}\oplus_3 T_{d_2}^{\{1\}}\})\label{14b}
		\\&=I(W_{[3]/ d_3};\{T_{d_3}^{\{1\}}\oplus_3 S_{\{1,3\}},T_{d_3}^{\{2\}}\oplus_3 S_{\{1,3\}}, T_{d_3}^{\{3\}}\oplus_3 S_{\{1,3\}} \}\notag
		\\&\qquad\cup\{T_{d_1}^{\{3\}}\oplus_3 S_{\{1,3\}},T_{d_2}^{\{3\}}\oplus_3 S_{\{1,3\}} \}\cup\{T_{d_1}^{\{2\}}\oplus_3 T_{d_2}^{\{1\}}\})\label{14c}
		\\&=I(W_{[3]/ d_3};\{W_{d_3}^1 \oplus_3 S_{d_3}\oplus_3 S_{\{1,3\}},W_{d_3}^2 \oplus_3 S_{d_3}\oplus_3 S_{\{1,3\}}, S_{d_3}\oplus_3 S_{\{1,3\}} \}\notag
		\\&\qquad\cup\{S_{d_1}\oplus_3 S_{\{1,3\}},S_{d_2}\oplus_3 S_{\{1,3\}} \}\cup\{W_{d_1}^2 \oplus_3 S_{d_1}\oplus_3 W_{d_2}^1 \oplus_3 S_{d_2}\})\label{14d}
		\\&=I(W_{[3]/ d_3};\{W_{d_3}^1 \oplus_3 S'_{d_3},W_{d_3}^2 \oplus_3 S'_{d_3}, S'_{d_3} \}\notag
		\\&\qquad\cup\{S'_{d_1},S'_{d_2}\}\cup\{W_{d_1}^2 \oplus_3 S'_{d_1}\oplus_3 W_{d_2}^1 \oplus_3 S'_{d_2}  \ominus_3 2S_{\{1,3\}}\})\label{14e}
		\\&=I(W_{[3]/ d_3};\{W_{d_3}^1 ,W_{d_3}^2, S'_{d_3} \}\cup\{S'_{d_1},S'_{d_2}\}\cup\{W_{d_1}^2 \oplus_3 W_{d_2}^1 \ominus_3 2S_{\{1,3\}}\})\label{14f}
		\\&=0,\label{secure22a}
	\end{align}
\end{subequations}
where (\ref{14a}) follows from the fact  that the last user can decode $\{T_{d_3}^{\{1\}}\oplus_3 S_{\{1,3\}},T_{d_3}^{\{2\}}\oplus_3 S_{\{1,3\}}, T_{d_3}^{\{3\}}\oplus_3 S_{\{1,3\}} \}$,  as discussed in the correctness condition,  (\ref{14b}) holds because
\begin{align*}
	H(2S_{\{1,3\}}\oplus_3 T_{d_1}^{\{3\}}\oplus_3 T_{d_3}^{\{1\}}\mid T_{d_3}^{\{1\}}\oplus_3 S_{\{1,3\}},T_{d_1}^{\{3\}}\oplus_3 S_{\{1,3\}})=0,\\
	H(2S_{\{2,3\}}\oplus_3 T_{d_2}^{\{3\}}\oplus_3 T_{d_3}^{\{2\}}\mid T_{d_3}^{\{2\}}\oplus_3 S_{\{1,3\}},T_{d_2}^{\{3\}}\oplus_3 S_{\{1,3\}}, S_{\{2,3\}}\ominus_3 S_{\{1,3\}} )=0,
\end{align*}
(\ref{14c}) holds because the random key $S_{\{2,3\}}$ is generated independently of everything else, (\ref{14d}) follows from the definition of the three shares in (\ref{Nan241224a}), (\ref{14e}) holds because we perform the following variable substitution
\begin{align}\label{15}
  S_{i} \oplus_3 S_{\{1,3\}} \triangleq S'_{i} , \quad \forall i\in[3],	
\end{align}
(\ref{14f}) holds by removing duplicates in a manner similar to how it is done for (\ref{14b}), and (\ref{secure22a}) holds because the random key $S_{\{1,3\}}$ is generated independently of everything else, and $S_{\{1,3\}}$ is also independent of $S'_{i}$ in  (\ref{15}).  Therefore,  the security condition (\ref{secureness}) can be satisfied.

\subsubsection{General Achievable Scheme for arbitrary $N$ and $K$} \hfill \break
\indent 
 The finite field $\mathbb{F}_{q}$ is taken to be sufficient large, i.e., $q$ is taken to be sufficient large. For $\forall K\ge 3,t \in\{1, \ldots, K-2\}$, we first generate shares which satisfy the \((\binom{K-1}{t-1}, \binom{K}{t})\) non-standard secret sharing property and then utilize these shares to built the proposed achievable scheme. 
 
 The shares are generated as follows. For all $k \in [K]$, divide \(W_k\) into $\binom{K}{t} - \binom{K-1}{t-1}$ equal parts, denoted as \(\{W_k^i,i\in[\binom{K}{t} - \binom{K-1}{t-1}]\}\), each of size \(\frac{F}{\binom{K}{t} - \binom{K-1}{t-1}}\), and generate $\binom{K-1}{t-1}$ random keys \(\{S_{k}^i,i\in[\binom{K-1}{t-1}]\}\) of size \(\frac{F}{\binom{K}{t} - \binom{K-1}{t-1}}\). The corresponding \(\binom{K}{t}\) shares of the \(k\)-th file \(W_k\) are denoted as \(\{T_{k}^{\mathcal{L}},\mathcal{L} \subset [K],|\mathcal{L}| = t\}\). Let 
\begin{align}
	\mathbf{v}_k&=\begin{bmatrix}W_k^1\ \cdots\ W_k^{\binom{K}{t} - \binom{K-1}{t-1}}\ S_{k}^1 \ \cdots \ S_{k}^{\binom{K-1}{t-1}}\end{bmatrix}^T,\nonumber
	\\\mathbf{T}_k&=\begin{bmatrix}T_k^{\mathcal{L}_1}\ \cdots\ T_k^{\mathcal{L}_{\binom{K}{t}}}\end{bmatrix}^T,\nonumber
\end{align}
where \(T_k^{\mathcal{L}_i}\) denotes the \(i\)-th share and the specific way in which the shares are sorted is not important in this context, and $\mathbf{T}_k$ can be defined by the  matrix $\mathbf{G}_{k}$ as follows,
\begin{align}
	\mathbf{T}_k&=\mathbf{G}_{k} \mathbf{v}_k \notag
	\\&=\begin{bmatrix}
		\begin{array}{c: c}
			\mathbf{I}_{(\binom{K}{t} - \binom{K-1}{t-1})\times(\binom{K}{t} - \binom{K-1}{t-1})} & \multirow{2}{*}{$\mathbf{V}_{\binom{K}{t}\times\binom{K-1}{t-1}}$} \\
			\cdashline{1-1}
			\mathbf{0}_{\binom{K-1}{t-1}\times( \binom{K}{t} - \binom{K-1}{t-1})} & 
		\end{array}
	\end{bmatrix}\mathbf{v}_k, \label{share}
\end{align}
where $\mathbf{0}_{\binom{K-1}{t-1}\times( \binom{K}{t} - \binom{K-1}{t-1})}$ represents the all-zero matrix with dimension $\binom{K-1}{t-1}\times( \binom{K}{t} - \binom{K-1}{t-1})$, $\mathbf{V}_{\binom{K}{t}\times\binom{K-1}{t-1}}$ represents a Vandermonde matrix with dimension $\binom{K}{t}\times\binom{K-1}{t-1}$, i.e., 
\begin{align}\label{17}
	\mathbf{V}_{\binom{K}{t}\times\binom{K-1}{t-1}}=\begin{bmatrix}
		1 & x_1 & x_1^2 & \cdots & x_1^{\binom{K-1}{t-1}-1} \\
		1 & x_2 & x_2^2 & \cdots & x_2^{\binom{K-1}{t-1}-1} \\
		\vdots & \vdots & \vdots & \ddots & \vdots \\
		1 & x_{\binom{K}{t}} & x_{\binom{K}{t}}^2 & \cdots & x_{\binom{K}{t}}^{\binom{K-1}{t-1}-1}
	\end{bmatrix},
\end{align}
where $\{x_i,i\in[{\binom{K}{t}}]\}$ are $\binom{K}{t}$ distinct elements from the finite field $\mathbb{F}_q$. The fact that the $\binom{K}{t}$ shares \(\{T_{k}^{\mathcal{L}},\mathcal{L} \subset [K],|\mathcal{L}| = t\}\) satisfy the $(\binom{K-1}{t-1}, \binom{K}{t})$ non-standard secret sharing property for $W_k$ is proved in Appendix \ref{app1}, i.e., any $\binom{K-1}{t-1}$ shares will not reveal any information about the secret $W_k$, but all $\binom{K}{t}$ shares will fully reveal $W_k$. 

In addition, the server needs to generate $\binom{K}{t+1}$ additional random keys, each of size $\frac{F}{\binom{K}{t} - \binom{K-1}{t-1}}$,  denoted as $\{S_\mathcal{V},\mathcal{V}\subset[K],|\mathcal{V}|=t+1\}$. Therefore, the total size of the randomness $LF$ generated by the server is $\frac{\binom{K-1}{t-1}F}{\binom{K}{t} - \binom{K-1}{t-1}} + \frac{\binom{K}{t+1}F}{\binom{K}{t} - \binom{K-1}{t-1}}$.

The proposed achievable scheme is built on the shares $\{T_{k}^{\mathcal{L}}\}$ defined in (\ref{share}), and the fact that for a given $k \in [K]$, the $\binom{K}{t}$ shares satisfy the $(\binom{K-1}{t-1}, \binom{K}{t})$ non-standard secret sharing property for $W_k$, will be used in the proof of the correctness condition (\ref{decode}) and the security condition (\ref{secureness}). 
More specifically, during the placement phase, for the first \(t+1\) caches, i.e., $Z_k,k\in[t+1]$, the cached content is
\begin{align*}
	&Z_k=\{T_n^\mathcal{L},\mathcal{L}\subset[K],|\mathcal{L}|=t,k\in \mathcal{L},n\in[N]\}\cup\{S_\mathcal{V},\mathcal{V}\subset[K],|\mathcal{V}|=t+1,\mathcal{V}\ne [t+1],k\in \mathcal{V} \}.
\end{align*}
For the last \(K-t-1\) caches, i.e., $Z_k,k\in[t+2:K]$, the cached content is
\begin{align*}
	Z_k&=\{T_n^\mathcal{L}\oplus_q S_{[t]\cup \{k\}},\mathcal{L}\subset[K],|\mathcal{L}|=t,k\in \mathcal{L},n\in[N]\}\\&\qquad\cup\{S_\mathcal{V}\ominus_q S_{[t]\cup \{k\}},\mathcal{V}\subset[K],|\mathcal{V}|=t+1,\mathcal{V}\ne [t]\cup\{k\},k\in \mathcal{V}\}.
\end{align*}

During the delivery phase, for a demand $(d_1,\ldots,d_k)$ where not all users request the same file, server delivers 
\begin{align*}
	&X_{(d_1,\cdots,d_K)}=\{\oplus_{i\in[t+1]}T_{d_i}^{[t+1]/\{i\}}\}\cup\{(t+1)S_\mathcal{V}\oplus_{i\in \mathcal{V}}T_{d_i}^{\mathcal{V}/\{i\}},\mathcal{V}\subset[K],|\mathcal{V}|=t+1,\mathcal{V}\ne[t+1]\}.
\end{align*}
Now that we have described the caching function and encoding function of the proposed scheme, we show that the correctness condition (\ref{decode}) and the security condition (\ref{secureness}) are satisfied in Appendix \ref{app4}.

 \begin{Remark}
 The reason why our proposed achievable scheme is built on a \emph{non-standard} secret sharing scheme rather than the Shamir secret sharing scheme is explained as follows. The similarity between the $(m,n)$  non-standard secret sharing scheme and the $(m,n)$ Shamir secret sharing scheme\cite{shamir1979share} lies in the fact that for both schemes, when the number of shares is less than $m$, no information about the secret is leaked. The difference, however, is that Shamir secret sharing requires the secret to be fully reconstructed once the number of shares exceeds the $m$. In contrast, non-standard secret sharing only requires the full reconstruction of the secret when all $n$ shares are collected. For cases where the number of shares lies between $m$ and $n$, no additional constraints are imposed. Intuitively, this relaxation of the security requirement leads to a share size reduction. From the perspective of implementation, the Shamir secret sharing scheme can be designed by using a Vandermonde matrix to linearly combine the entire file with random keys, resulting in each share being the size of one file size. In contrast, the non-standard secret sharing scheme can be designed by first dividing the file into smaller subfiles with size of $\frac{1}{n-m}$ file size and then using a Vandermonde matrix to linearly combine the subfiles with random keys. As a result, each share has the same size as a subfile, i.e., $\frac{1}{n-m}$ times the file size.
 \end{Remark}

\begin{Remark}
Compared to the general scheme in\cite{PCC2018}, which is also built on a non-standard secret sharing scheme, the improvement in our proposed scheme is that each user can cache one fewer random key and maintain the same delivery rate  while still satisfying the correctness and security conditions. More specifically, the first \(t+1\) users do not need to cache \(S_{[t+1]}\), and the subsequent \(K-t-1\) users with cached content  \(Z_k\) do not need to cache \(S_{[t] \cup \{k\}}\), which result in a reduction in cache size.
\end{Remark}

\section{Converse Proofs}\label{section5}
In this section, we provide the converse proofs of Theorems 1, 2 and 4. For Theorems  1 and 2, we first derive some properties that any  secure coded caching scheme must satisfy. Then, based on these properties, we derive the converse results. For Theorem 4, we mainly utilize the symmetry property in the coded caching problem in the converse proof.
\subsection{ Converse proof of Theorem 1} \label{Nan0127c}

When $M=1$, which means that the cache size is the same as the file size, we find that any secure coded caching scheme must satisfy the property that the cache content at the $k$-th user is a deterministic function of the delivery signal and the message it requests. We call this property the \emph{deterministic property of the cache content}. More specifically, we have the following lemma. 

{\textit{Lemma 1 
(deterministic property of the cache content):} } For $\forall N,K\ge 2$ and $M=1$, let $\mathcal{D}_s$ be the set of users who demand the same file as the $s$-th user, i.e., $\mathcal{D}_s=\{k \mid d_k=d_s\}$. When $|\mathcal{D}_s| \in [K-1]$, we have 
\begin{align}\label{lemma1}
H\left(W_{d_s},X_{(d_1,\cdots, d_K)}\right)=H\left(W_{d_s},Z_{\mathcal{D}_s},X_{(d_1,\cdots, d_K)} \right).
\end{align}		

\begin{IEEEproof} The proof of Lemma 1 is shown in Appendix \ref{app55}.
\end{IEEEproof}

Using the result of Lemma 1, we can derive another property that any secure coded caching scheme must satisfy when $M=1$, i.e., the mutual independence between any single file and the cached contents of the users. We call this the \emph{independence property between a file and cache contents}. More specifically, we have the following lemma. 

\textit{Lemma 2 
(independence property between a file and cache contents):} For $\forall N,K\ge 2$,  when $M=1$, we have that
 any single file $W_s$ and cache contents $Z_1,\ldots,Z_K$ are all mutually independent, i.e., 
\begin{align}\label{lemma2}
H(W_s,Z_{[1:K]})=H(W_s)+\sum_{i=1}^{K}H(Z_i).
\end{align}

\noindent \begin{IEEEproof}The proof of Lemma 2 is shown in Appendix \ref{app66}.\end{IEEEproof}

	Finally, armed with Lemmas 1 and 2, we are ready to prove the converse of Theorem 1.
 For $a \neq 1$, 
	we have
	\begin{subequations}\label{proof1}
		\begin{align}
			&H\left(X_{(1,\cdots,1,1,2)}\right)+H\left(X_{(1,\cdots,1,a,1)}\right)+2H(W_1)\notag\\
			&\ge  H(W_1,X_{(1,\cdots,1,1,2)})+H(W_1,X_{(1,\cdots,1,a,1)})\notag\\
			&= H(W_1,Z_{[K-1]},X_{(1,\cdots,1,1,2)})+H(W_1,Z_{[1:K-2]}, Z_K,X_{(1,\cdots,1,a,1)})\label{7a}\\  
			&\ge H(W_1,Z_{[1:K-2]}) +H(W_1,Z_{[1:K]},X_{(1,\cdots,1,1,2)},X_{(1,\cdots,1,a,1)})\label{7b}\\
			&\ge H(W_1,Z_{[1:K-2]})+H(W_1,W_2,W_a,X_{(1,\cdots,1,a,1)})\label{7c}\\
			&= H(W_1)+\sum_{i=1}^{K-2}H(Z_{i})+H(W_1,W_2,W_a)+H(X_{(1,\cdots,1,a,1)})\label{7d},
		\end{align}
	\end{subequations}
	where (\ref{7a}) holds due to Lemma 1, i.e., (\ref{lemma1}), (\ref{7b}) holds due to the submodular property of the entropy function,  i.e.,
	\begin{align}
		H(X_{\mathcal{A}})+H(X_{\mathcal{B}}) \geq H(X_{\mathcal{A} \bigcap \mathcal{B}})+H(X_{\mathcal{A} \bigcup \mathcal{B}}),  \label{submodular}
	\end{align}(\ref{7c}) follows from the correctness condition (\ref{decode}) and conditioning reduces entropy, and (\ref{7d}) follows from Lemma 2, i.e., (\ref{lemma2}), and the security condition (\ref{secureness}), i.e., any delivery signal can not reveal any information about any files except for the demand where all users request the same file. 
	
	From (\ref{7d}), we have
	\begin{align}\label{proof2}
		&H(X_{(1,\cdots,1,1,2)})\ge (K-3)H(W_1)+H(W_{\{1,2,a\}}).
	\end{align}
	If $N=2$, choose $a=2$ in (\ref{proof2}), and we have that the worst-case delivery rate $R$ satisfies
	\begin{align*}
		&RF\ge H(X_{(1,\cdots,1,1,2)})\ge (K-1)F.
	\end{align*}
	If $N\ge 3$, choose $a=3$ in (\ref{proof2}), and we have that the worst-case delivery rate $R$ satisfies
	\begin{align*}
		RF\ge H(X_{(1,\cdots,1,1,2)})\ge KF.
	\end{align*}
	Hence, the converse proof of Theorem 1 is complete.

\subsection{ Converse proof of Theorem 2} 
\label{Nan0127d}

Before presenting the converse results, we define the following notations. For $a \in [N], t \in [K]$, define $\mathcal{X}_{(t,a)}$ as the set of delivery signals where the $t$-th user requests file $W_a$, i.e., $\mathcal{X}_{(t,a)}=\{X_{(d_1,\cdots, d_K)}|d_t=a\}$. Further, let $\widetilde{ X}_{(t,a)}$ be an arbitrarily chosen element in the set $\mathcal{X}_{(t,a)}$, and define $\widetilde{ \mathcal{X}}_{(t,\mathcal{S})}$ as the set of the arbitrarily chosen elements, one from each  $\mathcal{X}_{(t,s)}$, $s \in \mathcal{S}$, i.e., $\widetilde{ \mathcal{X}}_{(t,\mathcal{S})}=\{{\widetilde{ X}_{(t,s)}|s\in \mathcal{S}}\}$.

When $R=1$, which means that the size of the delivery signal is the same as the file size, we find that any secure coded caching scheme must satisfy the property that the delivery signal  is a deterministic function of the cache content at the $k$-th user and the message it requests. We call this property the \emph{deterministic property of the delivery signal}. More specifically, we have the following lemma.
 
\textit{Lemma 3 (deterministic property of the delivery signal):} For any $ N ,K\ge 2$, and $R=1$, we have
\begin{align}\label{lemma3}
H(W_a,Z_t)=H(W_a,Z_t,\mathcal{X}_{(t,a)}).
\end{align}
\begin{IEEEproof}The proof of Lemma 3 is shown in Appendix \ref{app77}.\end{IEEEproof}

Using the result of Lemma 3, we can derive another property that any  secure coded caching scheme must satisfy when $R=1$, i.e., the mutual independence between any single file and a certain set of delivery signals. We call this the \emph{independence property between a file and delivery signals}. More specifically, we have the following lemma.

\textit{Lemma 4 
(independence property between a file and delivery signals):} For any $ N ,K\ge 2$, and $R=1$,  
for $\forall a,b \in [N]$, and $a\ne b$, we have
\begin{align}\label{lemma4}
	&H(W_b,\mathcal{X}_{(t,a)},\widetilde{\mathcal{X}}_{(t,[N]/\{a,b\})})=H(W_b) 
	+H(\mathcal{X}_{(t,a)})+\sum_{s\in [N]/\{a,b\}}H(\widetilde{X}_{(t,s)}). 
\end{align}	
Furthermore, we have
\begin{align}\label{cor1}
H(\mathcal{X}_{(t,a)},\widetilde{\mathcal{X}}_{(t,[N]/a)})=H(\mathcal{X}_{(t,a)})+\sum_{s\in [N]/a}H(\widetilde{X}_{(t,s)}).
\end{align}

\noindent\begin{IEEEproof}The proof of Lemma 4 is shown in Appendix \ref{app88}.\end{IEEEproof}
	
Finally, armed with Lemmas 3 and 4, we are ready to prove the converse of Theorem 2. We have
	\begin{subequations}
	\begin{align}
		\sum_{i=1}^{2}(H(W_1)+H(Z_i))&=\sum_{i=1}^{2}H(W_1,Z_i)\label{21a}
		\\&=\sum_{i=1}^{2}H(W_1,Z_i,\mathcal{X}_{(i,1)}) \label{21b}\\
		&\ge H\left(W_1,\mathcal{X}_{(1,1)} \bigcap \mathcal{X}_{(2,1)} \right)+H(W_{[N]},Z_1,Z_2,\mathcal{X}_{(1,1)},\mathcal{X}_{(2,1)}),\label{temp51}
	\end{align}
	\end{subequations}
where (\ref{21a}) holds due to the security condition (\ref{secureness}), (\ref{21b}) follows from Lemma 3, i.e., (\ref{lemma3}), and (\ref{temp51}) is due to the submodular property of the entropy function (\ref{submodular}), and the correctness condition (\ref{decode}). 
	
	Consider the two terms in (\ref{temp51}) separately. For the first term, we first define $K-2$ sets of delivery signals. More specifically, let 
	$\mathcal{G}_1$ be the set of $N-2$ delivery signals where $d_1=d_2=\cdots=d_{K-1}=1$, and $d_K \in [3:N]$. For $s \in [2:K-3]$, let $\mathcal{G}_s$ be the set of $N-1$ delivery signals where $d_1=d_2=\cdots=d_{K-s}=1$, $d_{K-s+2}=d_{K-s+3}=\cdots=d_K=2$, and $d_{K-s+1} \in \{1\} \bigcup [3:N]$. Let $\mathcal{G}_{K-2}$ be the set of $N$ delivery signals where $d_1=d_2=1$, $d_4=d_5=\cdots=d_K=2$, and $d_3 \in [1:N]$. 
	
	Thus, the first term in (\ref{temp51}) can be written as
	\begin{subequations}\label{temp52}
		\begin{align}
			H\left(W_1,\mathcal{X}_{(1,1)} \bigcap \mathcal{X}_{(2,1)}\right)
			&\ge H(W_1,\mathcal{G}_{[1:K-2]})\label{13a}\\
			&= H(W_1,\mathcal{G}_{[2:K-2]},\mathcal{G}_1)\notag\\
			&= H(\mathcal{G}_{[2:K-2]})+H(W_1)+\sum_{\widetilde{X}\in \mathcal{G}_{1}}H(\widetilde{X})\label{13b}\\
			&= H(\mathcal{G}_{[3:K-2]},\mathcal{G}_2)+H(W_1)+\sum_{\widetilde{X}\in \mathcal{G}_{1}}H(\widetilde{X})\notag\\
			&= H(\mathcal{G}_{[3:K-2]})+H(W_1)+\sum_{\widetilde{X}\in \mathcal{G}_{[1:2]}}H(\widetilde{X})\label{13c}\\
			&=\cdots= H(\mathcal{G}_{K-2})+H(W_1)+\sum_{\widetilde{X}\in \mathcal{G}_{[1:K-3]}}H(\widetilde{X})\label{13d}\\
			&= H(W_1)+\sum_{\widetilde{X}\in \mathcal{G}_{[1:K-2]}}H(\widetilde{X})\label{13e}\\
			&=((N-1)(K-2)+1)H(W_1)\label{13f},
		\end{align}
	\end{subequations}
	where (\ref{13a}) holds because $\mathcal{G}_{[1:K-2]}\subseteq (\mathcal{X}_{(1,1)} \bigcap \mathcal{X}_{(2,1)})$, i.e., all delivery signals in $\mathcal{G}_{[1:K-2]}$ satisfy that the first user and the second user both request $W_1$, (\ref{13b}) holds due to Lemma 4, i.e., (\ref{lemma4}) with $t=K$, $a=2$, $b=1$,
	(\ref{13c}) follows from (\ref{cor1}) with $t=K-1$, $a=2$, (\ref{13d}) follows from recursively using (\ref{cor1}) with $t=K-2,K-1, \cdots,4$, $a=2$, (\ref{13e}) follows from (\ref{cor1}) with $t=3$, $a=2$, and (\ref{13f}) holds because we are considering the case of $R=1$.
	
	For the second term in (\ref{temp51}), we have
	\begin{subequations}
	\begin{align}
		H(W_{[N]},Z_1,Z_2,\mathcal{X}_{(1,1)},\mathcal{X}_{(2,1)})&\ge H(W_{[N]},Z_1)\notag\\
		&= H(W_{[N]})+H(Z_1)\label{23a}\\
		&=NH(W_{1})+H(Z_1),\label{temp53}
	\end{align}
	\end{subequations}
	where (\ref{23a}) follows due to the security condition (\ref{secureness}).
	
	Finally, from (\ref{temp51}), (\ref{13f}) and (\ref{temp53}), we have that the cache size $M$ satisfies
	\begin{align*}
		&MF \ge H(Z_2)\ge (N-1)(K-1)F.
	\end{align*}
	Hence, the converse proof of Theorem 2 is complete.

\subsection{ Converse proof of Theorem 4}\label{section53}

According to the definitions of the cache size \( M \) and the worst-case delivery rate \( R \), 
\begin{align*}
	M\ge H(Z_2),\quad R\ge H(X_{1,\cdots,1,2}).
\end{align*}
To prove (\ref{theoeq4}), it is sufficient to prove
\begin{align}\label{fang11}
	(K-1)(K-2)H(Z_2)+2H(X_{1,\cdots,1,2})\ge K(K-1)H(W_1).
\end{align}
To prove (\ref{fang11}), it is sufficient to prove that the following four inequalities hold
\begin{align}
	&H(Z_2)+H(Z_1,X_{1,\cdots,1,2})\ge H(Z_{[2]},X_{1,\cdots,1,2})+H(W_1),\label{fang01}
	\\&H(X_{1,\cdots,1,2})+H(W_1,Z_{[2]})\ge H(Z_{\{1,K\}},X_{1,\cdots,1,2})+H(W_1),\label{fang02}
	\\&H(W_2,Z_{[2]},X_{1,\cdots,1,2})+H(Z_{\{1,K\}},X_{1,\cdots,1,2})\notag
	\\&\qquad\ge H(Z_{\{1,2,K\}},X_{1,\cdots,1,2})+H(Z_1,X_{1,\cdots,1,2})+H(W_1),\label{fang03}
	\\&(K-1)(K-2)H({Z_{[2]},X_{1,\cdots,1,2}})+2(K-2)H({Z_{\{1,2,K\}},X_{1,\cdots,1,2}})\notag
	\\&\qquad\ge(K-3)(K-2)H(Z_1,X_{1,\cdots,1,2})+2H(W_1,Z_{[2]})+2(K-2)H(W_2,Z_{[2]},X_{1,\cdots,1,2})\notag
	\\&\qquad\qquad+2(K-3)H(Z_{\{1,K\}},X_{1,\cdots,1,2}),  \label{fang04}
\end{align}
because if we multiply inequality (\ref{fang01}) by the coefficient $(K-1)(K-2)$, inequality (\ref{fang02}) by the coefficient $2$ and inequality (\ref{fang03}) by the coefficient $2(K-2)$, and then add the sum of the three resulting inequalities with (\ref{fang04}), we obtain (\ref{fang11}). 

The proofs of (\ref{fang01}), (\ref{fang02}), and (\ref{fang03}) can be found in Appendix \ref{app99}. In brief, the proofs primarily rely on the repeated application of the submodular property of the entropy function (\ref{submodular}), the correctness condition (\ref{decode}), and the security condition (\ref{secureness}). Additionally, the proof of (\ref{fang02}) involves utilizing the symmetry in user indexing, which will be defined in detail in Appendix \ref{app99}. Then, to prove (\ref{fang04}), it is sufficient to prove that the following three inequalities hold
\begin{align}
	&\frac{(K-1)(K-2)}{2}H(Z_{[2]},X_{1,\cdots,1,2})\notag
	\\&\qquad\ge \frac{(K-2)(K-3)}{2}H(Z_1,X_{1,\cdots,1,2})+\sum_{i=2}^{K-1}H(Z_{[i]},X_{1,\cdots,1,2}),\label{fanglem1fun1}
		\\&(K-2)H({Z_{\{1,2,K\}},X_{1,\cdots,1,2}})\ge (K-3)H({Z_{\{1,K\}},X_{1,\cdots,1,2}})+H(W_2,Z_{[2]},X_{1,\cdots,1,2}),\label{fanglem1fun2}
	\\&\sum_{i=2}^{K-1}H(Z_{[i]},X_{1,\cdots,1,2})\ge H(W_1,Z_{[2]})+(K-3)H(W_2,Z_{[2]},X_{1,\cdots,1,2})\label{fanglem1fun3},
\end{align}
because if we add the three inequalities together, i.e., (\ref{fanglem1fun1}),  (\ref{fanglem1fun2}) and  (\ref{fanglem1fun3}), and multiply the resulting inequality  by the coefficient $2$, we obtain (\ref{fang04}). 
The proof of (\ref{fanglem1fun1}), (\ref{fanglem1fun2}), (\ref{fanglem1fun3}) can be found in Appendix \ref{app99}. In brief, the proofs primarily rely on the repeated application of the submodular property of the entropy function (\ref{submodular}), the correctness condition (\ref{decode}), the security condition (\ref{secureness}) and the symmetry in user indexing, which will be defined in detail in Appendix \ref{app99}. Hence, the converse proof of Theorem 4 is complete.
\section{Conclusion}\label{section6}
	In this paper, we study the secure coded caching problem with $N$ files and $K$ users. In terms of achievability results, we propose three new schemes, one for the case of cache size \( M = 1 \), \( N = 2 \) and arbitrary \( K \), one for the case of rate \( R = 1 \) and arbitrary \( N, K \), and one for the case of cache size \( M > 1 \) and \( R > 1 \) and arbitrary \( N, K \). Compared to existing results, our schemes achieve a lower delivery rate under the same cache size.   Regarding the converse, we characterize new properties of the secure coded caching scheme for the end-points  of the memory-rate tradeoff curve. We further exploit the symmetry property of optimal secure coded caching schemes to derive new lower bounds. As a result, we characterize the two end-points of the optimal memory-rate tradeoff curve for arbitrary file number \( N \) and user number \( K \), as well as a segment of the curve where the cache size is relatively small for the case of two files  and arbitrary user number \( K \).

\appendices

\section{Proof of correctness and security conditions for the proposed scheme in Section \ref{Nan0127a}} \label{app2}
The correctness condition (\ref{decode}) can be rewritten as
\begin{align}
	H(W_{d_k}\mid Z_k, X_{(d_1, \cdots, d_K)})
	&=H(Z_k, X_{(d_1, \cdots, d_K)}\mid W_{d_k})+H\left(W_{d_k}\right) -H(Z_k, X_{(d_1, \cdots, d_K)})\notag\\
	&=\left(r\left(\begin{bmatrix}\mathbf{G}^T_{Z_k}\ \mathbf{G}^T_{X}\end{bmatrix}^T (\mathbf{I}-\mathbf{E}_{d_k})\right)+1-r\left(\begin{bmatrix}\mathbf{G}^T_{Z_k} \ \mathbf{G}^T_{X}\end{bmatrix}^T \right) \right) F. \label{rank1}
\end{align}

The security condition (\ref{secureness}) can be rewritten as
\begin{subequations}
\begin{align}
	I(W_{[N] /\left\{d_{k}\right\}} ;  Z_{k},X)
	&=H(Z_{k},X)-H(Z_{k},X |W_{[N] /\{d_{k}\}})\notag\\
	&=H(Z_{k},X)-H(Z_{k},X |W_{\bar{d}_k})\label{Nan0127e}\\
	&=\left( r(\begin{bmatrix}\mathbf{G}^T_{Z_k}\ \mathbf{G}^T_{X}\end{bmatrix}^T)-r(\begin{bmatrix}\mathbf{G}^T_{Z_k}\ \mathbf{G}^T_{X}\end{bmatrix}^T (\mathbf{I}-\mathbf{E}_{\bar{d}_k})) \right) F.\label{rank2}
\end{align}
\end{subequations}
where (\ref{Nan0127e}) follows because we consider the case of two files. 

\begin{Remark}\label{remark3}
	$\mathbf{G}(\mathbf{I}-\mathbf{E}_{d_i})$ is a matrix that is the same as $\mathbf{G}$, except that the $d_i$-th column is set to the all-zero vector. 
\end{Remark}

For $\forall k\in [K-1]$, we have
\begin{align}
r(\begin{bmatrix}\mathbf{G}^T_{Z_k}\ \mathbf{G}^T_{X}\end{bmatrix}^T)
	&=r\left(\left[
	\begin{array}{cc:ccc}
		0 & 0&&\mathbf{e}^T_k& \\
		\hdashline
		\delta_{1d_1}a_1 & \delta_{2d_1}a_1&2&& \\
		\multicolumn{2}{c:}{\cdots}&&\ddots \\
		\delta_{1d_{K-1}}a_{K-1} & \delta_{2d_{K-1}}a_{K-1}&&&2\\
	\end{array}\right]	\right)\notag\\
	&=r\left(\left[
	\begin{array}{cc:ccc}
		\delta_{1d_k}a_k & \delta_{2d_k}a_k&0&\cdots& 0\\
		\hdashline
		\delta_{1d_1}a_1 & \delta_{2d_1}a_1&2&& \\
		\multicolumn{2}{c:}{\cdots}&&\ddots \\
		\delta_{1d_{K-1}}a_{K-1} & \delta_{2d_{K-1}}a_{K-1}&&&2\\
	\end{array}\right]	\right)\label{35},
\end{align}
and for the $K$-th user, we have
\begin{align}
	r(\begin{bmatrix} \mathbf{G}^T_{Z_K} \ \mathbf{G}^T_{X}\end{bmatrix}^T)
	&=r\left(\left[
	\begin{array}{cc:ccc}
		2 & 2&1&\cdots&1\\ 
		\hdashline
		\delta_{1d_1}a_1 & \delta_{2d_1}a_1&2&& \\
		\multicolumn{2}{c:}{\cdots}&&\ddots \\
		\delta_{1d_{K-1}}a_{K-1} & \delta_{2d_{K-1}}a_{K-1}&&&2\\
	\end{array}\right]	\right)\notag\\
	&=r\left(\left[
	\begin{array}{cc:ccc}
		\delta_{1d_K}a_K & \delta_{2d_K}a_K&0&\cdots& 0\\ 
		\hdashline
		\delta_{1d_1}a_1 & \delta_{2d_1}a_1&2&& \\
		\multicolumn{2}{c:}{\cdots}&&\ddots \\
		\delta_{1d_{K-1}}a_{K-1} & \delta_{2d_{K-1}}a_{K-1}&&&2\\
	\end{array}\right]	\right)\label{36},
\end{align}
where (\ref{35}) and (\ref{36}) follow from using elementary row transformations and  (\ref{matrixconstrain}). Hence, it can be summarized that for $\forall k\in [K]$, we have
\begin{align*}
	r(\begin{bmatrix}\mathbf{G}^T_{Z_k} \ \mathbf{G}^T_{X}\end{bmatrix}^T)
	&=r\left(\left[
	\begin{array}{cc:ccc}
		\delta_{1d_k}a_k & \delta_{2d_k}a_k&0&\cdots& 0\\
		\hdashline
		\delta_{1d_1}a_1 & \delta_{2d_1}a_1&2&& \\
		\multicolumn{2}{c:}{\cdots}&&\ddots \\
		\delta_{1d_{K-1}}a_{K-1} & \delta_{2d_{K-1}}a_{K-1}&&&2\\
	\end{array}\right]	\right).
\end{align*}
When $d_k=1$, due to (\ref{Nan0127f}), we have
\begin{align}
	r(\begin{bmatrix}\mathbf{G}^T_{Z_k} \ \mathbf{G}^T_{X}\end{bmatrix}^T)
	&=r\left(\left[
	\begin{array}{cc:ccc}
		a_k & 0&0&\cdots& 0\\
		\hdashline
		\delta_{1d_1}a_1 & \delta_{2d_1}a_1 &2&& \\
		\multicolumn{2}{c:}{\cdots}&&\ddots \\
		\delta_{1d_{K-1}}a_{K-1} & \delta_{2d_{K-1}}a_{K-1}&&&2\\
	\end{array}\right]	\right)=K.\label{rGG}
\end{align}
Furthermore, according to Remark \ref{remark3}, we have
\begin{align}
	&r(\begin{bmatrix}\mathbf{G}^T_{Z_k}\ \mathbf{G}^T_{X}\end{bmatrix}^T (\mathbf{I}-\mathbf{E}_{d_k}))=K-1, \label{rGGcorrect}\\
	&r(\begin{bmatrix}\mathbf{G}^T_{Z_k}\ \mathbf{G}^T_{X}\end{bmatrix}^T (\mathbf{I}-\mathbf{E}_{\bar{d_k}}))=K. \label{rGGsecure}
\end{align}
When $d_k=2$, (\ref{rGG}), (\ref{rGGcorrect}), (\ref{rGGsecure}) still hold for the same reason. Thus, following from (\ref{rank1}) and (\ref{rank2}), we have that the correctness and security conditions are both satisfied.

\section{Proof of correctness and security conditions for the proposed scheme in Section \ref{Nan0127b}} \label{app3}
The correctness condition (\ref{decode}) can be rewritten to be the same as (\ref{rank1}), and the security condition (\ref{secureness}) can be rewritten as
\begin{align}
	I(W_{[N] /\left\{d_{k}\right\}} ;  Z_{k},X)
	&=H(Z_{k},X)-H(Z_{k},X |W_{[N] /\{d_{k}\}})\notag\\
	&=\Bigg(r(\begin{bmatrix}\mathbf{G}^T_{Z_k}\ \mathbf{G}^T_{X}\end{bmatrix}^T) -r\left(\begin{bmatrix}\mathbf{G}^T_{Z_k}\ \mathbf{G}^T_{X}\end{bmatrix}^T (\mathbf{I}-\sum_{i \in [N] /\{d_{k}\}} \mathbf{E}_{d_i}) \right) \Bigg)F. \label{Nan0127h}
\end{align}
\begin{Remark}\label{Nan0127g}
	$\mathbf{G}(\mathbf{I}-\sum_{i\in\mathcal{A} }\mathbf{E}_{d_i})$ is a matrix that is the same as $\mathbf{G}$, except that the $d_i$-th column is set to the all-zero vector for all $i\in \mathcal{A}$. 
\end{Remark}	

When $d_i\ne 1$ and $k \in [K-1]$, to calculate the rank of matrices involved in (\ref{rank1}) and (\ref{Nan0127h}), we have
\begin{subequations}
\begin{align}
	r(\begin{bmatrix}\mathbf{G}^T_{Z_k} \ \mathbf{G}^T_{X}\end{bmatrix}^T)
	&=r(\begin{bmatrix}  \mathbf{G}^T_{X} \ \mathbf{G}^T_{Z_k} \end{bmatrix}^T)\notag\\
	&=r(\begin{bmatrix}  \mathbf{G}^T_{X} \ \bar{\mathbf{G}}^T_{Z_k} \end{bmatrix}^T)\label{39a}\\
	&=r\left(\left[
	\begin{array}{c:c}
		\mathbf{e}_{d_K}^T&\textstyle\bigoplus_{i=1}^{K}(\mathbf{e}^T_{d_K-1,i}\oplus \mathbf{e}^T_{d_i-1,i})\\
		\hdashline
		\mathbf{e}_{1}^T\oplus\mathbf{e}_{2}^T&\mathbf{e}^T_{1k}\\
		\mathbf{e}_{1}^T\oplus\mathbf{e}_{3}^T& \mathbf{e}^T_{2k}\\
		\cdots& \cdots\\
		\mathbf{e}_{1}^T\oplus\mathbf{e}_{N}^T& \mathbf{e}^T_{N-1k}\\
		\hdashline
		\mathbf{0}&\mathbf{I}'
	\end{array}\right]\right)\notag\\
	&=r\left(\left[
	\begin{array}{c:c}
		\mathbf{e}_{d_K}^T&\mathbf{e}^T_{d_K-1,k}\oplus \mathbf{e}^T_{d_k-1,k}\\
		\hdashline
		\mathbf{e}_{1}^T\oplus\mathbf{e}_{2}^T&\mathbf{e}^T_{1k}\\
		\mathbf{e}_{1}^T\oplus\mathbf{e}_{3}^T& \mathbf{e}^T_{2k}\\
		\cdots& \cdots\\
		\mathbf{e}_{1}^T\oplus\mathbf{e}_{N}^T& \mathbf{e}^T_{N-1k}\\
		\hdashline
		\mathbf{0}&\mathbf{I}'
	\end{array}\right]\right)\label{39b}\\
	&=r\left(\left[
	\begin{array}{c:c}
		\mathbf{e}_{d_K}^T\oplus\mathbf{e}_{1}^T\oplus\mathbf{e}_{d_K}^T\oplus\mathbf{e}_{1}^T\oplus\mathbf{e}_{d_i}^T&\mathbf{0}\\
		\hdashline
		\mathbf{e}_{1}^T\oplus\mathbf{e}_{2}^T&\mathbf{e}^T_{1k}\\
		\mathbf{e}_{1}^T\oplus\mathbf{e}_{3}^T& \mathbf{e}^T_{2k}\\
		\cdots& \cdots\\
		\mathbf{e}_{1}^T\oplus\mathbf{e}_{N}^T& \mathbf{e}^T_{N-1k}\\
		\hdashline
		\mathbf{0}&\mathbf{I}'
	\end{array}\right]\right)\label{39c}\\
	&=r\left(\left[
	\begin{array}{c:c}
		\mathbf{e}_{d_i}^T&\mathbf{0}\\
		\hdashline
		\mathbf{e}_{1}^T\oplus\mathbf{e}_{2}^T&\mathbf{e}^T_{1k}\\
		\mathbf{e}_{1}^T\oplus\mathbf{e}_{3}^T& \mathbf{e}^T_{2k}\\
		\cdots& \cdots\\
		\mathbf{e}_{1}^T\oplus\mathbf{e}_{N}^T& \mathbf{e}^T_{N-1k}\\
		\hdashline
		\mathbf{0}&\mathbf{I}'
	\end{array}\right]\right)\label{rankGX2},
\end{align}
\end{subequations} 
where (\ref{39a}) follows due to the definition of $\bar{\mathbf{G}}_{Z_k}$, which is the same as $\mathbf{G}_{Z_k}$ except for some rearranging of the order of rows. More specifically, for the matrix $\mathbf{G}_{Z_k}$, $k \in [K-1]$, there are $N-1$ rows that combine random keys with files, i.e., the first $N$ columns of these $N-1$ rows are not all 0. We extract these $N-1$ rows of $\mathbf{G}_{Z_k}$ and place them as the top rows in $\bar{\mathbf{G}}_{Z_k}$, and the remaining part of $\mathbf{G}_{Z_k}$ can be denoted as $\begin{bmatrix} \mathbf{0} \ \mathbf{I}' \end{bmatrix}$, and are placed as the bottom rows in $\bar{\mathbf{G}}_{Z_k}$, and (\ref{39b}), (\ref{39c}) both follow from performing elementary row transformations.  

It can be verified that equation (\ref{rankGX2}) still holds for $d_i = 1$, $k \in [K-1]$. Hence, for $k \in [K-1]$, we have
\begin{align}
	&r(\begin{bmatrix}\mathbf{G}^T_{Z_k} \ \mathbf{G}^T_{X}\end{bmatrix}^T)=(N-1)(K-1)+1.\nonumber
\end{align}
According to Remarks \ref{remark3} and \ref{Nan0127g}, we have
\begin{align*}
	r(\begin{bmatrix}\mathbf{G}^T_{Z_k}\ \mathbf{G}^T_{X}\end{bmatrix}^T (\mathbf{I}-\mathbf{E}_{d_k}))&=(N-1)(K-1),\ \text{and} \\
	r(\begin{bmatrix}\mathbf{G}^T_{Z_k}\ \mathbf{G}^T_{X}\end{bmatrix}^T (\mathbf{I}-\sum_{i\in[N] /\{d_{k}\}}\mathbf{E}_{d_i}))&=(N-1)(K-1)+1,
\end{align*}
which means that the correctness condition (\ref{rank1})  and the security condition (\ref{Nan0127h}) are both satisfied. 

Furthermore, the correctness and security conditions are also satisfied for the $K$-th user by repeating the procedure above.

\section{Proof of correctness and security conditions for the proposed scheme in Section \ref{section43}} \label{app4}
The correctness condition  (\ref{decode}) can be satisfied for the following reasons.  For $\forall k\in[t+1]$, the $k$-th user with cached content $Z_k$ can use $\{T_{d_i}^{[t+1]/\{i\}},i\in[t+1],i\ne k\}$ to decode  $T_{d_k}^{[t+1]/\{k\}}$ from $\oplus_{i\in[t+1]}T_{d_i}^{[t+1]/\{i\}}\in X_{(d_1,d_2,d_3)}$, and use $\{T_{d_i}^{\mathcal{V}/\{i\}},S_\mathcal{V},\mathcal{V}\subset[K],|\mathcal{V}|=t+1,\mathcal{V}\ne[t+1],k\in \mathcal{V},i\in \mathcal{V}/\{k\}\}$  to decode $\{T_{d_k}^{\mathcal{V}/\{k\}},\mathcal{V}\subset[K],|\mathcal{V}|=t+1,\mathcal{V}\ne[t+1]\}$ from $\{(t+1)S_\mathcal{V}\oplus_{i\in \mathcal{V}}T_{d_i}^{\mathcal{V}/\{i\}},\mathcal{V}\subset[K],|\mathcal{V}|=t+1,\mathcal{V}\ne[t+1]\}$, because for $\mathcal{V}\subset[K],|\mathcal{V}|=t+1,\mathcal{V}\ne[t+1]$, we have
\begin{align*}
	&((t+1)S_\mathcal{V}\oplus_{i\in \mathcal{V}}T_{d_i}^{\mathcal{V}/\{i\}})\ominus_q
	(\oplus_{i\in[\mathcal{V}]/\{k\}}T_{d_i}^{\mathcal{V}/\{i\}}\oplus_q (t+1)S_\mathcal{V})
	=T_{d_k}^{\mathcal{V}/\{k\}}.
\end{align*}
Thus, the $k$-th user can decode \(\{T_{d_k}^{{\mathcal{V}}/\{k\}}, |\mathcal{V}|=t+1, \mathcal{V}\subset[K], k\in \mathcal{V}\}\). Combined with the shares \(\{T_{d_k}^{\mathcal{L}}, |\mathcal{L}|=t, \mathcal{L}\subset[K], k\in \mathcal{L}\}\) cached during the placement phase, the $k$-th user can obtain all shares of \(W_{d_k}\), thus fully decoding \(W_{d_k}\).

For $\forall k\in[t+2:K]$,  the $k$-th user with cached content $Z_k$ can use $\{T_{d_i}^{\mathcal{V}/\{i\}}\oplus_q S_{[t]\cup \{k\}},S_\mathcal{V}\ominus_q S_{[t]\cup \{k\}},\mathcal{V}\subset[K],|\mathcal{V}|=t+1,k\in \mathcal{V},i\in \mathcal{V}/\{k\}\}$ to decode $\{T_{d_k}^{\mathcal{V}/\{k\}}\oplus_q S_{[t]\cup\{k\}},\mathcal{V}\subset[K],|\mathcal{V}|=t+1,k\in \mathcal{V}\}$ from $\{(t+1)S_\mathcal{V}\oplus_{i\in \mathcal{V}}T_{d_i}^{\mathcal{V}/\{i\}},\mathcal{V}\subset[K],|\mathcal{V}|=t+1,k\in \mathcal{V}\}$, because for $\mathcal{V}\subset[K],|\mathcal{V}|=t+1,k\in \mathcal{V}$, we have
\begin{align*}
	&((t+1)S_\mathcal{V}\oplus_{i\in \mathcal{V}}T_{d_i}^{\mathcal{V}/\{i\}})\notag
	\\&\qquad\ominus_q
	(\oplus_{i\in[\mathcal{V}]/\{k\}}((T_{d_i}^{\mathcal{V}/\{i\}}\oplus_q S_{[t]\cup \{k\}})\oplus_q (S_\mathcal{V}\ominus_q S_{[t]\cup \{k\}})))
	\ominus_q (S_\mathcal{V}\ominus_q S_{[t]\cup \{k\}})
	\\&=((t+1)S_\mathcal{V}\oplus_{i\in \mathcal{V}}T_{d_i}^{\mathcal{V}/\{i\}})\ominus_q(\oplus_{i\in[\mathcal{V}]/\{k\}}(T_{d_i}^{\mathcal{V}/\{i\}}\oplus_q S_\mathcal{V}))	\ominus_q (S_\mathcal{V}\ominus_q S_{[t]\cup \{k\}})
	\\&=((t+1)S_\mathcal{V}\oplus_{i\in \mathcal{V}/\{k\}}T_{d_i}^{\mathcal{V}/\{i\}}\oplus_q T_{d_k}^{\mathcal{V}/\{k\}})
	\ominus_q(\oplus_{i\in[\mathcal{V}]/\{k\}}T_{d_i}^{\mathcal{V}/\{i\}}\oplus_q tS_\mathcal{V})\ominus_q (S_\mathcal{V}\ominus_q S_{[t]\cup \{k\}})
	\\&=T_{d_k}^{\mathcal{V}/\{k\}}\oplus_q S_{[t]\cup \{k\}}.
\end{align*}
Thus, the $k$-th user can decode \(\{T_{d_k}^{\mathcal{V}/\{k\}}\oplus_q S_{[t]\cup \{k\}}, |\mathcal{V}|=t+1, \mathcal{V}\subset[K], k\in \mathcal{V}\}\). Combined with the shares \(\{T_{d_k}^{\mathcal{L}}\oplus_q S_{[t]\cup \{k\}}, |\mathcal{L}|=t, \mathcal{L}\subset[K], k\in \mathcal{L}\}\) cached during the placement phase, the \(k\)-th user can obtain all shares of \(W_{d_k}\) modulo a random key \(S_{[t] \cup \{k\}}\), i.e., $\{T_{d_k}^{\mathcal{L}}\oplus_q S_{[t]\cup \{k\}}, |\mathcal{L}|=t, \mathcal{L}\subset[K]\}$. Next, we prove that since the shares are defined according to (\ref{share}),  \(W_{d_k}\) can be decoded using these modulo sums. Note that $S_{[t] \cup \{k\}}$ is a random key of size \(\frac{F}{\binom{K}{t} - \binom{K-1}{t-1}}\) and let 
\begin{align}
	\mathbf{v}_{k,S_{[t] \cup \{k\}}}&=\begin{bmatrix}W_k^1\ \cdots\ W_k^{\binom{K}{t} - \binom{K-1}{t-1}}\ S_{[t] \cup \{k\}} \ S_{k}^1 \ \cdots \ S_{k}^{\binom{K-1}{t-1}}\end{bmatrix}^T,\nonumber
	\\\mathbf{T}_{k,S_{[t] \cup \{k\}}}&=\begin{bmatrix}T_k^{\mathcal{L}_1}\oplus_q S_{[t] \cup \{k\}}\ \cdots\ T_k^{\mathcal{L}_{\binom{K}{t}}}\oplus_q S_{[t] \cup \{k\}}\end{bmatrix}^T\nonumber,
\end{align}
where $\mathbf{T}_{k,S_{[t] \cup \{k\}}}$ can be represented by the  matrix $\mathbf{G}_{k,S_{[t] \cup \{k\}}}$ as follows,
\begin{align*}
	\mathbf{T}_{k,S_{[t] \cup \{k\}}}&=\mathbf{G}_{k,S_{[t] \cup \{k\}}} \mathbf{v}_{k,S_{[t] \cup \{k\}}},
\end{align*}
where 
\begin{align*}
	\mathbf{G}_{k,S_{[t] \cup \{k\}}}=
	\begin{bmatrix}
		\begin{array}{c: c:c}
			\mathbf{I}_{(\binom{K}{t} - \binom{K-1}{t-1})\times(\binom{K}{t} - \binom{K-1}{t-1})} & \multirow{2}{*}{$\mathbf{1}_{\binom{K}{t}\times 1}$}&\multirow{2}{*}{$\mathbf{V}_{\binom{K}{t}\times\binom{K-1}{t-1}}$}  \\
			\cdashline{1-1}
			\mathbf{0}_{\binom{K-1}{t-1}\times( \binom{K}{t} - \binom{K-1}{t-1})} & &
		\end{array}
	\end{bmatrix},
\end{align*}
where $\mathbf{1}_{\binom{K}{t}\times1}$ represents the all-one matrix with dimension $\binom{K}{t}\times1$. Considering that the first column of the Vandermonde matrix \(\mathbf{V}_{\binom{K}{t} \times \binom{K-1}{t-1}}\) is also a column vector of all ones, we treat \( S_{[t] \cup \{k\}} \oplus_q S_{k}^1 \) as a random key denoted by \( S' \). Let
\begin{align*}
	\mathbf{v}_{k,S'} = \begin{bmatrix} W_k^1 \ \cdots \ W_k^{\binom{K}{t} - \binom{K-1}{t-1}} \ S' \ S_{k}^2 \ \cdots \ S_{k}^{\binom{K-1}{t-1}}\end{bmatrix}^T,
\end{align*}
then \(\mathbf{T}_{k,S_{[t] \cup \{k\}}}\) can be represented by \(\mathbf{v}_{k,S'}\) and \(\mathbf{G}_{k,S'}\) as follows
\begin{align*}
	\mathbf{T}_{k,S_{[t] \cup \{k\}}}&=\mathbf{G}_{k,S'} \mathbf{v}_{k,S'},
\end{align*}
where 
\begin{align*}
	\mathbf{G}_{k,S'}	&=\begin{bmatrix}
		\begin{array}{c: c}
			\mathbf{I}_{(\binom{K}{t} - \binom{K-1}{t-1})\times(\binom{K}{t} - \binom{K-1}{t-1})}  &\multirow{2}{*}{$\mathbf{V}_{\binom{K}{t}\times\binom{K-1}{t-1}}$}  \\
			\cdashline{1-1}
			\mathbf{0}_{\binom{K-1}{t-1}\times( \binom{K}{t} - \binom{K-1}{t-1})} & 
		\end{array}
	\end{bmatrix}
	\\&=\mathbf{G}_{k},
\end{align*}
i.e., by considering the modular sum of two random keys as a single random key, the resulting  matrix $\mathbf{G}_{k,S'}$ is the same as the  matrix $\mathbf{G}_{k}$ defined in (\ref{share}). Therefore, it follows that after obtaining all elements of \(\mathbf{T}_{k,S_{[t] \cup \{k\}}}\),   the subfiles \( W_k^1, \cdots, W_k^{\binom{K}{t} - \binom{K-1}{t-1}} \) can be decoded, thereby revealing \( W_k \). Thus, we have completed the proof of the correctness condition (\ref{decode}).

The security condition (\ref{secureness}) can be satisfied for the following reasons. For the first $t+1$ users, considering the $k$-th user with cached content $Z_k$, $k\in[t+1]$, we have
\begin{subequations}
	\begin{align}
		&I(W_{[N]/ d_k};Z_k,X_{(d_1,\cdots,d_K)})\notag
		\\&=I(W_{[N]/ d_k};\{T_n^\mathcal{L},\mathcal{L}\subset[K],|\mathcal{L}|=t,k\in \mathcal{L},n\in[N]\}\notag
		\\&\qquad\cup\{S_\mathcal{V},\mathcal{V}\subset[K],|\mathcal{V}|=t+1,\mathcal{V}\ne [t+1],k\in \mathcal{V} \}\cup\{\oplus_{i\in[t+1]}T_{d_i}^{[t+1]/\{i\}}\}\notag
		\\&\qquad\cup\{(t+1)S_\mathcal{V}\oplus_{i\in \mathcal{V}}T_{d_i}^{\mathcal{V}/\{i\}},\mathcal{V}\subset[K],|\mathcal{V}|=t+1\notag
		,\mathcal{V}\ne[t+1]\})\notag
		\\&= I(W_{[N]/ d_k}; \{T_{d_k}^{\mathcal{L}'},\mathcal{L}'\subset[K],|\mathcal{L}'|=t\}\cup\{T_n^\mathcal{L},\mathcal{L}\subset[K],|\mathcal{L}|=t,k\in \mathcal{L},n\in[N]/ \{d_k\}\}\notag
		\\&\qquad\cup\{S_\mathcal{V},\mathcal{V}\subset[K],|\mathcal{V}|=t+1,\mathcal{V}\ne [t+1],k\in \mathcal{V} \}\cup\{\oplus_{i\in[t+1]}T_{d_i}^{[t+1]/\{i\}}\}\notag
		\\&\qquad\cup\{(t+1)S_\mathcal{V}\oplus_{i\in \mathcal{V}}T_{d_i}^{\mathcal{V}/\{i\}},\mathcal{V}\subset[K],|\mathcal{V}|=t+1,\mathcal{V}\ne[t+1]\})\label{secure1a}
		\\&= I(W_{[N]/ d_k};\{T_{d_k}^{\mathcal{L}'},\mathcal{L}'\subset[K],|\mathcal{L}'|=t\}\cup \{T_n^\mathcal{L},\mathcal{L}\subset[K],|\mathcal{L}|=t,k\in \mathcal{L},n\in[N]/ \{d_k\}\}\notag
		\\&\qquad\cup\{S_\mathcal{V},\mathcal{V}\subset[K],|\mathcal{V}|=t+1,\mathcal{V}\ne [t+1],k\in \mathcal{V} \}\notag
		\\&\qquad\cup\{(t+1)S_\mathcal{V}\oplus_{i\in \mathcal{V}}T_{d_i}^{\mathcal{V}/\{i\}},\mathcal{V}\subset[K],|\mathcal{V}|=t+1,k\notin \mathcal{V},\mathcal{V}\ne[t+1]\})\label{secure1b}
		\\&= I(W_{[N]/ d_k};\{T_{d_k}^{\mathcal{L}'},\mathcal{L}'\subset[K],|\mathcal{L}'|=t\}\cup \{T_n^\mathcal{L},\mathcal{L}\subset[K],|\mathcal{L}|=t,k\in \mathcal{L},n\in[N]/ \{d_k\}\})\label{secure1c}
		\\&=0,\label{secure1d}
 	\end{align}
\end{subequations}
where (\ref{secure1a}) holds because it has already been shown in the correctness condition proof that all shares corresponding to \(W_{d_k}\) can be decoded, (\ref{secure1b}) holds by removing duplicate elements from the sets. Intuitively, these removed terms are composed of the sum of a share that a user needs and the shares and random keys the user already has, and when the requested shares, i.e., $\{T_{d_k}^{\mathcal{L}'},\mathcal{L}'\subset[K],|\mathcal{L}'|=t\}$, are isolated in (\ref{secure1a}), these terms can be eliminated, (\ref{secure1c}) holds because the random keys $\{S_\mathcal{V}\}$ are generated independently of everything else, and  (\ref{secure1d}) holds due to the fact that for $n\in[N]/ \{d_k\}$, the shares  $\{T_n^\mathcal{L},\mathcal{L}\subset[K],|\mathcal{L}|=t,k\in \mathcal{L}\}$ satisfy the non-standard secret sharing property for $W_{n}$. 

For the last $K-t-1$ users, considering the $k$-th user with cached content $Z_k$, $k\in[t+2:K]$, we have
\begin{subequations}
	\begin{align}
		&I(W_{[N]/ d_k};Z_k,X_{(d_1,\cdots,d_K)})\notag
		\\&=I(W_{[N]/ d_k};\{T_n^\mathcal{L}\oplus_q S_{[t]\cup \{k\}},\mathcal{L}\subset[K],|\mathcal{L}|=t,k\in \mathcal{L},n\in[N]\}\notag
		\\&\qquad\cup\{S_\mathcal{V}\ominus_q S_{[t]\cup \{k\}},\mathcal{V}\subset[K],|\mathcal{V}|=t+1,\mathcal{V}\ne [t]\cup\{k\},k\in \mathcal{V}\}\notag\cup\{\oplus_{i\in[t+1]}T_{d_i}^{[t+1]/\{i\}}\}\notag
		\\&\qquad\cup\{(t+1)S_\mathcal{V}\oplus_{i\in \mathcal{V}}T_{d_i}^{\mathcal{V}/\{i\}},\mathcal{V}\subset[K],|\mathcal{V}|=t+1,\mathcal{V}\ne[t+1]\})\notag 
		\\&=I(W_{[N]/ d_k};\{T_{d_k}^{\mathcal{L}'}\oplus_q S_{[t]\cup \{k\}},\mathcal{L}'\subset[K],|\mathcal{L}'|=t\}\notag
		\\&\qquad\cup\{T_n^\mathcal{L}\oplus_q S_{[t]\cup \{k\}},\mathcal{L}\subset[K],|\mathcal{L}|=t,k\in \mathcal{L},n\in[N]/ \{d_k\}\}\notag
		\\&\qquad\cup\{S_\mathcal{V}\ominus_q S_{[t]\cup \{k\}},\mathcal{V}\subset[K],|\mathcal{V}|=t+1,\mathcal{V}\ne [t]\cup\{k\},k\in \mathcal{V}\}\cup\{\oplus_{i\in[t+1]}T_{d_i}^{[t+1]/\{i\}}\}\notag
		\\&\qquad\cup\{(t+1)S_\mathcal{V}\oplus_{i\in \mathcal{V}}T_{d_i}^{\mathcal{V}/\{i\}},\mathcal{V}\subset[K],|\mathcal{V}|=t+1,\mathcal{V}\ne[t+1]\})\label{secure2a}
		\\&=I(W_{[N]/ d_k};\{T_{d_k}^{\mathcal{L}'}\oplus_q S_{[t]\cup \{k\}},\mathcal{L}'\subset[K],|\mathcal{L}'|=t\}\notag
		\\&\qquad\cup\{T_n^\mathcal{L}\oplus_q S_{[t]\cup \{k\}},\mathcal{L}\subset[K],|\mathcal{L}|=t,k\in \mathcal{L},n\in[N]/ \{d_k\}\}\notag
		\\&\qquad\cup\{S_\mathcal{V}\ominus_q S_{[t]\cup \{k\}},\mathcal{V}\subset[K],|\mathcal{V}|=t+1,\mathcal{V}\ne [t]\cup\{k\},k\in \mathcal{V}\}\cup\{\oplus_{i\in[t+1]}T_{d_i}^{[t+1]/\{i\}}\}\notag
		\\&\qquad\cup\{(t+1)S_\mathcal{V}\oplus_{i\in \mathcal{V}}T_{d_i}^{\mathcal{V}/\{i\}},\mathcal{V}\subset[K],|\mathcal{V}|=t+1,k\notin \mathcal{V},\mathcal{V}\ne[t+1]\})\label{secure2b}
		\\&=I(W_{[N]/ d_k};\{T_{d_k}^{\mathcal{L}'}\oplus_q S_{[t]\cup \{k\}},\mathcal{L}'\subset[K],|\mathcal{L}'|=t\}\notag
		\\&\qquad\cup\{T_n^\mathcal{L}\oplus_q S_{[t]\cup \{k\}},\mathcal{L}\subset[K],|\mathcal{L}|=t,k\in \mathcal{L},n\in[N]/ \{d_k\}\}\cup\{\oplus_{i\in[t+1]}T_{d_i}^{[t+1]/\{i\}}\}),\label{secure2c}
	\end{align} 
\end{subequations}
where (\ref{secure2a}) holds because it has already been shown in the correctness condition proof that all shares of their respective  \(W_{d_k}\) each combined with a random key modulo operation, i.e., $\{T_{d_k}^{\mathcal{L}}\oplus_q S_{[t]\cup \{k\}}, |\mathcal{L}|=t, \mathcal{L}\subset[K]\}$, can be decoded, (\ref{secure2b}) holds by removing duplicate elements from the sets, (\ref{secure2c}) holds because the random keys $\{S_\mathcal{V}\}$ are generated independently of everything else, similarly to (\ref{secure1b}). Here, we use a similar variable substitution technique used in the correctness condition proof, i.e., for $\forall n\in[N]$, let the random key corresponding to the first column of the Vandermonde matrix \(\mathbf{V}_{\binom{K}{t} \times \binom{K-1}{t-1}}\) defined in (\ref{17}) be denoted as \(S_n^1\) in the shares of the \(n\)-th file, where all elements of the first column are 1. The sum of this key \(S_{n}^1\)  with \(S_{[t] \cup \{k\}}\) can be regarded as a single random key, so we define 
	\begin{align*}
		\tilde{T}_n^\mathcal{L} = T_n^\mathcal{L} \oplus_q S_{[t] \cup \{k\}}, \quad  \forall n\in[N].
	\end{align*}
 According to the proof of the correctness condition, for $\forall n\in[N]$, the shares \(\{\tilde{T}_{n}^\mathcal{L},\mathcal{L}\subset[K],|\mathcal{L}|=t\}\) satisfy the \((\binom{K-1}{t-1}, \binom{K}{t})\) non-standard secret sharing property for \(W_n\).
Therefore, we have (\ref{secure2c}) is equal to
\begin{subequations}
	\begin{align}
		&I(W_{[N]/ d_k};\{T_{d_k}^{\mathcal{L}'}\oplus_q S_{[t]\cup \{k\}},\mathcal{L}'\subset[K],|\mathcal{L}'|=t\}\notag
		\\&\qquad\cup\{T_n^\mathcal{L}\oplus_q S_{[t]\cup \{k\}},\mathcal{L}\subset[K],|\mathcal{L}|=t,k\in \mathcal{L},n\in[N]/ \{d_k\}\}\cup\{\oplus_{i\in[t+1]}T_{d_i}^{[t+1]/\{i\}}\})\notag
		\\&=I(W_{[N]/ d_k};\{\tilde{T}_{d_k}^{\mathcal{L}'},\mathcal{L}'\subset[K],|\mathcal{L}'|=t\}\notag
		\\&\qquad\cup\{\tilde{T}_n^\mathcal{L},\mathcal{L}\subset[K],|\mathcal{L}|=t,k\in \mathcal{L},n\in[N]/ \{d_k\}\}\cup\{\oplus_{i\in[t+1]}(\tilde{T}_{d_i}^{[t+1]/\{i\}}\ominus_q S_{[t]\cup \{k\}})\})\notag
		\\&=I(W_{[N]/ d_k};\{\tilde{T}_{d_k}^{\mathcal{L}'},\mathcal{L}'\subset[K],|\mathcal{L}'|=t\}\notag
		\\&\qquad\cup\{\tilde{T}_n^\mathcal{L},\mathcal{L}\subset[K],|\mathcal{L}|=t,k\in \mathcal{L},n\in[N]/ \{d_k\}\}\notag
		\\&\qquad\cup\{\oplus_{i\in[t+1]}(\tilde{T}_{d_i}^{[t+1]/\{i\}})\ominus_q (t+1)_{\text{mod } q}S_{[t]\cup \{k\}}\})\label{secure3c}
		\\&=I(W_{[N]/ d_k};\{\tilde{T}_{d_k}^{\mathcal{L}'},\mathcal{L}'\subset[K],|\mathcal{L}'|=t\}\cup\{\tilde{T}_n^\mathcal{L},\mathcal{L}\subset[K],|\mathcal{L}|=t,k\in \mathcal{L},n\in[N]/ \{d_k\}\})\label{secure3a}
		\\&=0,\label{secure3b}
	\end{align}
\end{subequations}
where $(t+1)_{\text{mod } q}$ in (\ref{secure3c}) represents the result of \( t+1 \) modulo \( q \). Since \( q \) is chosen to be sufficiently large in this context, it can be ensured that \( (t+1)_{\text{mod } q} \neq 0 \),  (\ref{secure3a}) holds because \(S_{[t] \cup \{k\}}\) is independent of \(\{\tilde{T}_{n}^\mathcal{L},\mathcal{L}\subset[K],|\mathcal{L}|=t,n\in[N]\} \cup\{W_{[N]}\} \).  To be more specific, all the elements in \( \{\tilde{T}_n^\mathcal{L}\} \) contain an XOR operation with \( S_n^1 \oplus_q S_{[t] \cup \{k\}} \) and from the construction of shares \( \{{T}_n^\mathcal{L}\} \) in (\ref{share}), where the first column elements of the Vandermonde matrix \(\mathbf{V}_{\binom{K}{t} \times \binom{K-1}{t-1}}\) in (\ref{17}) are all 1, it can be concluded that every \( {T}_n^\mathcal{L} \) contains and only contains a single $S_n^1$. Since both \( S_n^1 \) and \( S_{[t] \cup \{k\}} \) are generated independently of everything else, from \(\{\tilde{T}_{n}^\mathcal{L},\mathcal{L}\subset[K],|\mathcal{L}|=t,n\in[N]\} \cup\{W_{[N]}\} \), it is always impossible to decode \( S_{[t] \cup \{k\}} \) alone which demonstrates the independence between \(S_{[t] \cup \{k\}}\) and \(\{\tilde{T}_{n}^\mathcal{L},\mathcal{L}\subset[K],|\mathcal{L}|=t,n\in[N]\} \cup\{W_{[N]}\} \). Finally,  (\ref{secure3b}) holds because for $n\in[N]/ \{d_k\}$, the shares $\{\tilde{T}_n^\mathcal{L},\mathcal{L}\subset[K],|\mathcal{L}|=t,k\in \mathcal{L}\}$ satisfy the non-standard secret sharing property for $W_{n}$. Thus, the proof of correctness and security conditions is complete.
 
 \begin{Remark}
 	It is worth mentioning that since the shares corresponding to different files are generated independently, if the shares for each individual file do not leak any information about its corresponding file, then the shares of these different files together will not leak any information about these files either. In other words, the shares of different files are independent from each other.
 \end{Remark}

\section{Proof of non-standard secret sharing property for the proposed shares in Section \ref{section43}}\label{app1}
In this appendix, we verify that the shares we construct in (\ref{share}) satisfy the $(\binom{K-1}{t-1}, \binom{K}{t})$ non-standard secret sharing property, i.e., 1) any number of shares less than the threshold \(\binom{K-1}{t-1}\) do not leak any information about the file; and 2) all shares combined  fully reconstruct the file. 

For the \(k\)-th file \(W_k\), to verify 1), consider arbitrarily selecting $\binom{K-1}{t-1}$ rows from $\mathbf{G}_{k}$ to form a matrix $\mathbf{G}'_{k}$ of dimensions $\binom{K-1}{t-1} \times \binom{K}{t}$. i.e.,
\begin{align*}
	\mathbf{G}'_{k}=\begin{bmatrix}
		\mathbf{B}_{\binom{K-1}{t-1}\times(\binom{K}{t}-\binom{K-1}{t-1})}\	\mathbf{\mathbf{V}}'_{\binom{K-1}{t-1} \times \binom{K-1}{t-1}}
	\end{bmatrix},
\end{align*}
where the matrix \(\mathbf{B}_{\binom{K-1}{t-1}\times(\binom{K}{t}-\binom{K-1}{t-1})}\) represents the matrix that corresponds to the multiplication of the subfile elements of \(W_k\). The last $\binom{K-1}{t-1}$ columns of this matrix form a $\binom{K-1}{t-1} \times \binom{K-1}{t-1}$ matrix $\mathbf{\mathbf{V}}'_{\binom{K-1}{t-1} \times \binom{K-1}{t-1}}$, which remains a Vandermonde matrix and its row rank and column rank are both \(\binom{K-1}{t-1}\). Since the matrix \(\mathbf{G}'_{k}\) has \(\binom{K-1}{t-1}\) rows, it follows that the matrix \(\mathbf{G}'_{k}\) is full row rank. Let us name the shares corresponding to  matrix $\mathbf{G}'_{k}$ as $\mathbf{T}'_k$. Then we have:
\begin{align*}
	I\left(W_k ;\mathbf{T}'_k\right)
	&= H(\mathbf{T}'_k)-H(\mathbf{T}'_k|W_k)
	\\&= r(\mathbf{G}'_{k})-r\left(\begin{bmatrix} \mathbf{0}_{\binom{K-1}{t-1}\times(\binom{K}{t}-\binom{K-1}{t-1})}\	\mathbf{\mathbf{V}}'_{\binom{K-1}{t-1} \times \binom{K-1}{t-1}} \end{bmatrix} \right)
	\\&=0.
\end{align*}
From this, it follows that any $\binom{K-1}{t-1}$ shares do not leak information about the file $W_k$.

To verify 2), consider the matrix $\mathbf{G}''_{k}$ formed by the last $\binom{K-1}{t-1}$ rows of $\mathbf{G}_{k}$, i.e.,
\begin{align*}
	\mathbf{G}''_{k}=\begin{bmatrix}
		\mathbf{0}_{\binom{K-1}{t-1}\times(\binom{K}{t}-\binom{K-1}{t-1})}\	\mathbf{\mathbf{V}}''_{\binom{K-1}{t-1} \times \binom{K-1}{t-1}}
	\end{bmatrix}.
\end{align*}
Suppose the shares corresponding to this matrix $\mathbf{G}''_{k}$ are denoted as $\mathbf{T}''_k$. When $\mathbf{T}''_k$ is obtained, due to $\mathbf{V}''_{\binom{K-1}{t-1} \times \binom{K-1}{t-1}}$ being a full-rank Vandermonde matrix, $ S_{k}^1, \cdots, S_{k}^{\binom{K-1}{t-1}}$ can be solved from $\mathbf{T}''_k$. Consequently, upon acquiring the remaining shares, $\mathbf{G}_k$ can be utilized to solve all sub-files $W_k^1, \cdots, W_k^{\binom{K}{t} - \binom{K-1}{t-1}}$, thereby revealing $W_k$. Thus, the proof of non-standard secret sharing property for the proposed shares in Section \ref{section43} is complete.

\section{Proof of Lemma 1}\label{app55}
First, we show that 
\begin{align}\label{wx=wzx}
	H(W_{d_s},X_{(d_1,\cdots, d_K)})= H(W_{d_s},Z_s,X_{(d_1,\cdots, d_K)}) 
\end{align}
is true as follows,
\begin{subequations}
\begin{align}
	H(W_{d_s},X_{(d_1,\cdots, d_K)})&=H(W_{d_s})+H(X_{(d_1,\cdots, d_K)})\label{46a}\\
	& \geq H(Z_s)+H(X_{(d_1,\cdots, d_K)})\label{46b}
	\\&\ge H(Z_s,X_{(d_1,\cdots, d_K)})\notag \\
	&=H(W_{d_s},Z_s,X_{(d_1,\cdots, d_K)}) \label{Nan0123a}
	\\&\ge H(W_{d_s},X_{(d_1,\cdots, d_K)})\label{46d},
\end{align}
\end{subequations}
where (\ref{46a}) holds due to the security condition (\ref{secureness}), i.e., any delivery signal can not reveal any information about any files except for the demand where all users request the same file, (\ref{46b}) holds because we are considering the case of $M=1$, and (\ref{Nan0123a}) holds due to the correctness condition in (\ref{decode}). Since the beginning and end of (\ref{46d}) are the same, the inequalities are actually equalities in (\ref{Nan0123a}), and therefore we have (\ref{wx=wzx}). 

Now, we are ready to prove (\ref{lemma1}). 
Since the right-hand side (RHS) is no smaller than the left-hand side (LHS) of (\ref{lemma1}), we only need to prove that the LHS $\geq$ RHS. Towards this end, we have
\begin{subequations}
\begin{align}
	|\mathcal{D}_s|H(W_{d_s},X_{(d_1,\cdots, d_K)})&=\sum_{i\in\mathcal{D}_s}H(W_{d_s},Z_i,X_{(d_1,\cdots, d_K)})\label{47a}\\
	&\ge(|\mathcal{D}_s|-1)H(W_{d_s},X_{(d_1,\cdots, d_K)})+H(W_{d_s},Z_{\mathcal{D}_s},X_{(d_1,\cdots, d_K)}), \label{Nan0123b}
\end{align}
\end{subequations}
where (\ref{47a}) follows from (\ref{wx=wzx}), and (\ref{Nan0123b}) follows from $|\mathcal{D}_s|-1$ repeated use of the submodular property of the entropy function (\ref{submodular}).
Hence, (\ref{lemma1}) follows from (\ref{Nan0123b}), and Lemma 1 is proved. 

\section{Proof of Lemma 2}  \label{app66}
Without loss of generality, we may set $s=1$. We use the notation $X'_l$ to represent the delivery signal $X_{(d_1,\cdots, d_K)}$ when $d_1=\cdots=d_{K-l}=1,d_{K-l+1}=\cdots=d_K=2$, $l \in [K-1]$. We have
\begin{subequations}
\begin{align}
	H(W_1,Z_{[1:K]})+H(Z_1,X'_1)
	&=H(W_1,Z_{[1:K]})+H(W_1,Z_{1},X'_1)\label{48a}\\
	& \geq H(W_1,Z_{[1:K]})+H(W_1,X'_1)\notag\\
	&=H(W_1,Z_{[1:K]})+H(W_1,Z_{[1:K-1]},X'_1)\label{48b}\\
	&\ge H(W_1,Z_{[1:K-1]})+H(W_1,Z_{[1:K]},X'_1)\label{48c}. 
\end{align}
\end{subequations}
For the second term of (\ref{48c}), we further have
\begin{subequations}
\begin{align}
	H(W_1,Z_{[1:K]},X'_1)&=H(W_2,W_1,Z_{[1:K]},X'_1)\label{48d}\\
	&\ge H(W_2,Z_1,X'_1) \notag\\
	&= H(W_2)+H(Z_1,X'_1)\label{48e}\\
	&\geq H(Z_K)+H(Z_1,X'_1)\label{48f},
\end{align}
\end{subequations}
where (\ref{48a}) and (\ref{48d}) hold due to the correctness condition (\ref{decode}), (\ref{48b}) holds due to Lemma 1, i.e., (\ref{lemma1}), (\ref{48c}) follows from the submodular property of the entropy function (\ref{submodular}), (\ref{48e}) follows from the security condition (\ref{secureness}), and (\ref{48f}) follows from the condition that $M=1$. 

Combining (\ref{48c}) with (\ref{48f}), we have
\begin{align}
	H(W_1,Z_{[1:K]}) \geq H(W_1,Z_{[1:K-1]})+H(Z_K).\nonumber
\end{align}

Using the same steps above with $X'_2$ in place of $X'_1$, we have
\begin{align}
	H(W_1,Z_{[1:K-1]}) \geq H(W_1, Z_{[1:K-2]})+H(Z_{K-1}). \nonumber
\end{align}
Hence, by repeating the steps above by combining 
$H(W_1,Z_{[1:K-2]}), \cdots, H(W_1,Z_{[1:2]})$  with $X'_3,\cdots ,X'_{K-1}$, respectively, Lemma 2 can be 
proved.

\section{Proof of Lemma 3} \label{app77}
The idea of the proof is almost the same as Lemma 1. 
Let $\widetilde{ X}_{(t,a)}$ represent one element in the set $\mathcal{X}_{(t,a)}$,  i.e., $\widetilde{ X}_{(t,a)} \in \mathcal{X}_{(t,a)}$. 
Then, first, we show that 
\begin{align}
	H(W_{a},Z_t)= H(W_{a},Z_t,\widetilde{ X}_{(t,a)})\label{wz=wzx}
\end{align}
is true. This is because
\begin{subequations}
\begin{align}
	H(W_{a},Z_t)
	&=H(W_{a})+H(Z_t)\label{511a}\\
	& \geq H(\widetilde{ X}_{(t,a)})+H(Z_t)\label{511b}\\
	&\ge H(Z_t,\widetilde{ X}_{(t,a)})\notag\\
	&=H(W_{a},Z_t,\widetilde{ X}_{(t,a)})\label{511c}\\
	&\ge H(W_{a},Z_t),\label{Fang0123a}
\end{align}
\end{subequations}
where (\ref{511a}) holds due to the security condition (\ref{secureness}), i.e., any cache content can not reveal any information about any message, (\ref{511b}) holds because we are considering the case of $R=1$, and (\ref{511c}) holds due to the correctness condition in (\ref{decode}). Since the beginning and end of (\ref{Fang0123a}) are the same, the inequalities are actually equalities in (\ref{Fang0123a}), and we have (\ref{wz=wzx}). 

Now, we are ready to prove (\ref{lemma3}). 
Since the RHS is no smaller than the LHS of (\ref{lemma3}), we only need to prove that the LHS $\geq$ RHS. We have
\begin{subequations}
\begin{align}
	|\mathcal{X}_{(t,a)}|H(W_{a},Z_t)&=\sum_{\widetilde{X}\in \mathcal{X}_{(t,a)}}H(W_{a},Z_t,\widetilde{ X})\label{52a}\\
	&\ge(|\mathcal{X}_{(t,a)}|-1)H(W_{a},Z_t)+H(W_a,Z_t,\mathcal{X}_{(t,a)}),\label{Fang0123b}
\end{align}
\end{subequations}
where (\ref{52a}) follows from (\ref{wz=wzx}), and (\ref{Fang0123b}) follows from $|\mathcal{X}_{(t,a)}|-1$ repeated use of the submodular property of the entropy function (\ref{submodular}). Hence, (\ref{lemma3}) follows from (\ref{Fang0123b}), and Lemma 3 can be proved.

\section{Proof of Lemma 4} \label{app88}
First, we prove (\ref{lemma4}). Since the RHS is no smaller than the LHS of (\ref{lemma4}), we only need to prove that the LHS $\geq$ RHS. We have
\begin{subequations}
\begin{align}
	&H(W_b,\mathcal{X}_{(t,a)},\widetilde{\mathcal{X}}_{(t,[N]/\{a,b\})})+H(W_a,Z_t)\notag\\
	&=H(W_b,\mathcal{X}_{(t,a)},\widetilde{\mathcal{X}}_{(t,[N]/\{a,b\})})+H(W_a,Z_t,\mathcal{X}_{(t,a)})\label{53a}\\
	&\ge H(\mathcal{X}_{(t,a)})+H(W_{[N]},Z_{t},\mathcal{X}_{(t,a)},\widetilde{\mathcal{X}}_{(t,[N]/\{a,b\})})\label{53b}\\
	&\ge H(\mathcal{X}_{(t,a)})+H(W_{[N]},Z_{t},\widetilde{X}_{(t,a)})\notag\\
	&\ge H(\mathcal{X}_{(t,a)})+\sum_{s\in [N]/a}H(W_s)+H(W_{a},Z_{t},\widetilde{X}_{(t,a)})\label{53c}\\
	&\ge H(\mathcal{X}_{(t,a)})+H(W_b)+\sum_{s\in [N]/\{a,b\}}H(\widetilde{X}_{(t,s)})+H(W_{a},Z_{t},\widetilde{X}_{(t,a)})\label{53d}\\
	&\geq H(\mathcal{X}_{(t,a)})+H(W_b)+\sum_{s\in [N]/\{a,b\}}H(\widetilde{X}_{(t,s)})+H(W_{a},Z_{t})\notag,
\end{align}
\end{subequations}
where (\ref{53a}) holds due to Lemma 3, i.e., (\ref{lemma3}), (\ref{53b}) holds due to the submodular property of the entropy function (\ref{submodular}) and the correctness condition (\ref{decode}), where we have utilized the fact that $\widetilde{\mathcal{X}}_{(t,[N]/\{a,b\})}$ consists of delivery signals where the $t$-th user requests every file in $[N]/\{a,b\}$, (\ref{53c}) follows from the security condition (\ref{secureness}), and (\ref{53d}) holds because we are considering the case of $R=1$. This completes the proof of (\ref{lemma4}).

As for (\ref{cor1}), the proof follows the same steps as that of (\ref{lemma4}), except that we remove the terms of $H(W_b)$ and $W_b$, and replace $[N]/ \{a,b\}$ with $[N]/ \{a\}$.  Thus, the proof of Lemma 4 is complete. 

\section{Proof of intermediate inequalities in section \ref{section53}} \label{app99}
In this appendix, we will prove that the intermediate inequalities (\ref{fang01}), (\ref{fang02}), (\ref{fang03}), (\ref{fanglem1fun1}), (\ref{fanglem1fun2}), and (\ref{fanglem1fun2}) are true.

Before the proof, we first have the following proposition which shows that similar to the argument in \cite{tianchao}, it is sufficient to only consider user-index-symmetric schemes in the secure coded caching (SCC) problem. To define user-index-symmetric schemes, we assume a permutation function $\pi(\cdot)$ on the user index set of $[K]$. Denote  $\mathcal{W}=\left\{W_{1},\cdots, W_{N}\right\}$,  $\mathcal{Z}=\left\{Z_{1},\cdots, Z_{K}\right\}$, $\mathcal{X}=\left\{X_{d_{1}, \cdots, d_{K}}: d_{k} \in[N]\right\}$. Define
\begin{subequations}
	\begin{align}
		\pi(\mathcal{Z})&\triangleq\left\{\pi(Z_{1}),\cdots,\pi( Z_{K})\right\},\notag
		\\
		\pi(\mathcal{X})&\triangleq\left\{\pi(X_{d_{1}, \cdots, d_{K}}): d_{k} \in[N]\right\},\notag
	\end{align}
\end{subequations}
where $\pi(Z_{k})\triangleq Z_{\pi(k)},\pi(X_{d_1,\cdots,d_K})\triangleq X_{d_{\pi(1)},\cdots,d_{\pi(K)}}$.
We call an SCC scheme \emph{user-index-symmetric}\cite{tianchao}, if for any subsets $\mathcal{W}_{o_1}\subseteq \mathcal{W},\mathcal{Z}_{o_2}\subseteq \mathcal{Z},\mathcal{X}_{o_3}\subseteq \mathcal{X}$, and
any permutation $\pi(\cdot)$, we have
$H\left(\mathcal{W}_{{o_1}}, \mathcal{Z}_{{o_2}}, \mathcal{X}_{{o_3}}\right)=H\left(\mathcal{W}_{{o_1}}, \pi\left(\mathcal{Z}_{{o_2}}\right), \pi\left(\mathcal{X}_{{o_3}}\right)\right)$.
Next, we have the following proposition.

\textit{Proposition 1	(symmetry in  user indexing of SCC):}  For any SCC scheme, there always exists a corresponding  user-index-symmetric scheme,  whose worst-case delivery rate is  no larger than that of the original scheme.
\begin{IEEEproof} 
	The proof follows the same argument as \cite{tianchao} and is thus omitted.
\end{IEEEproof}

Proposition 1 tells us that when deriving converse results, we may focus on user-index-symmetric schemes only, and this is without loss of generality.  

Now, we are ready to prove (\ref{fang01}), (\ref{fang02}), (\ref{fang03}), (\ref{fanglem1fun1}), (\ref{fanglem1fun2}), and (\ref{fanglem1fun2}).
Firstly, (\ref{fang01}) holds because
\begin{subequations}
	\begin{align}
		H(Z_2)+H(Z_1,X_{1,\cdots,1,2})
		&= H(Z_2)+H(X_{1,\cdots,1,2})+H(Z_{1},X_{1,\cdots,1,2})-H(X_{1,\cdots,1,2})\notag
		\\&\ge H(Z_2,X_{1,\cdots,1,2})+H(Z_{1},X_{1,\cdots,1,2})-H(X_{1,\cdots,1,2})\label{fangsub1a}
		\\&= H(W_1,Z_2,X_{1,\cdots,1,2})+H(W_1,Z_{1},X_{1,\cdots,1,2})-H(X_{1,\cdots,1,2})\label{fangsub1b}
		\\&\ge H(W_1,X_{1,\cdots,1,2})+H(W_1,Z_{[2]},X_{1,\cdots,1,2})-H(X_{1,\cdots,1,2})\label{fangsub1c}
		\\&= H(W_1)+H(W_1,Z_{[2]},X_{1,\cdots,1,2}),\label{fangsub1d}
	\end{align}
\end{subequations}
where (\ref{fangsub1a}) and (\ref{fangsub1c}) holds due to the submodular property of the entropy function (\ref{submodular}), (\ref{fangsub1b}) holds due to the correctness condition (\ref{decode}), and (\ref{fangsub1d}) holds due to the security condition (\ref{secureness}).

Secondly, (\ref{fang02}) holds because
\begin{subequations}
	\begin{align}
		H(X_{1,\cdots,1,2})+H(W_1,Z_{[2]})&=H(X_{1,\cdots,1,2})+H(Z_1)+H(W_1,Z_{\{1,K\}})-H(Z_1)\label{fangsub2a}
		\\&\ge H(Z_1,X_{1,\cdots,1,2})+H(W_1,Z_{\{1,K\}})-H(Z_1)\label{fangsub2b}
		\\&= H(W_1,Z_1,X_{1,\cdots,1,2})+H(W_1,Z_{\{1,K\}})-H(Z_1)\label{fangsub2c}
		\\&\ge H(W_1,Z_{1})+H(Z_{\{1,K\}},W_1, X_{1,\cdots,1,2})-H(Z_1)\label{fangsub2d}
		\\&= H(Z_{\{1,K\}},W_1, X_{1,\cdots,1,2})+H(W_1),\label{fangsub2e}
		\\&= H(Z_{\{1,K\}},X_{1,\cdots,1,2})+H(W_1),\label{Nan241228b}
	\end{align}
\end{subequations}
where (\ref{fangsub2a}) holds due to the symmetry in user indexing, (\ref{fangsub2b}) and (\ref{fangsub2d}) hold due to the submodular property of the entropy function (\ref{submodular}), (\ref{fangsub2c}) and (\ref{Nan241228b})  holds due to the correctness condition (\ref{decode}), and (\ref{fangsub2e}) holds due to the security condition (\ref{secureness}).

Thirdly, (\ref{fang03}) holds because
\begin{subequations}
	\begin{align}
		&H(W_2,Z_{[2]},X_{1,\cdots,1,2})+H(Z_{\{1,K\}},X_{1,\cdots,1,2})\notag\\
		&=H(W_2,Z_{[2]},X_{1,\cdots,1,2})+H(W_2,Z_{\{1,K\}},X_{1,\cdots,1,2})\label{fangsub3a}
		\\&\ge H(W_2,Z_{1},X_{1,\cdots,1,2})+H(W_2,Z_{\{1,2,K\}},X_{1,\cdots,1,2})\label{fangsub3b}
		\\&= H(Z_1,X_{1,\cdots,1,2})+H(W_2, Z_{\{1,2,K\}},X_{1,\cdots,1,2})+H(W_2),\label{fangsub3c}
		\\&= H(Z_{\{1,2,K\}},X_{1,\cdots,1,2})+H(Z_1,X_{1,\cdots,1,2})+H(W_2),\label{Nan241228c}
		\\&= H(Z_{\{1,2,K\}},X_{1,\cdots,1,2})+H(Z_1,X_{1,\cdots,1,2})+H(W_1),\label{Nan241228d}
	\end{align}
\end{subequations}
where (\ref{fangsub3a}) and (\ref{Nan241228c}) holds due to the correctness condition (\ref{decode}), (\ref{fangsub3b}) holds due to the submodular property of the entropy function (\ref{submodular}), (\ref{fangsub3c}) holds due to the security condition (\ref{secureness}), and (\ref{Nan241228d}) follows because the entropy of the messages $W_1, \cdots, W_K$ are equal. 

Fourthly, (\ref{fanglem1fun1}) holds because
\begin{subequations}
	\begin{align}
		\frac{(K-1)(K-2)}{2}H(Z_{[2]},X_{1,\cdots,1,2})
		&=\sum_{i=2}^{K-1}\sum_{j=2}^{i}H(Z_{\{1,j\}},X_{1,\cdots,1,2})\label{fangsub4a}
		\\&\ge\sum_{i=2}^{K-1}((i-2)H(Z_{1},X_{1,\cdots,1,2})+H(Z_{[i]},X_{1,\cdots,1,2}))\label{fangsub4b}
		\\&= \frac{(K-2)(K-3)}{2}H(Z_1,X_{1,\cdots,1,2})+\sum_{i=2}^{K-1}H(Z_{[i]},X_{1,\cdots,1,2}), \notag
	\end{align}
\end{subequations}
where (\ref{fangsub4a}) holds due to the symmetry in user indexing, and (\ref{fangsub4b}) holds due to $i-2$ repeated use of the submodular property of the entropy function (\ref{submodular}).

Fifthly, (\ref{fanglem1fun2}) holds because
\begin{subequations}
	\begin{align}
		(K-2)H({Z_{\{1,2,K\}},X_{1,\cdots,1,2}})
		&=\sum_{i=2}^{K-1}H(Z_{\{1,i,K\}},X_{1,\cdots,1,2})\label{fangsub5a}
		\\&\ge(K-3)H(Z_{\{1,K\}},X_{1,\cdots,1,2})+H(Z_{[K]},X_{1,\cdots,1,2})\label{fangsub5b}
		\\&=(K-3)H(Z_{\{1,K\}},X_{1,\cdots,1,2})+H(W_2,Z_{[K]},X_{1,\cdots,1,2})\label{fangsub5c}
		\\&\ge(K-3)H(Z_{\{1,K\}},X_{1,\cdots,1,2})+H(W_2,Z_{[2]},X_{1,\cdots,1,2}),\notag
	\end{align}
\end{subequations}
where (\ref{fangsub5a}) holds due to the symmetry in user indexing, (\ref{fangsub5b}) holds due to $K-3$ repeated use of the submodular property of the entropy function (\ref{submodular}), and (\ref{fangsub5c}) holds due to the correctness condition (\ref{decode}).

Lastly, (\ref{fanglem1fun3}) holds because
\begin{subequations}
	\begin{align}
		&\sum_{i=2}^{K-1}H(Z_{[i]},X_{1,\cdots,1,2})\notag\\
		&=\sum _{i=2}^{K-2}H(Z_{[i]},X_{1,\cdots,1,2})+H(W_1,Z_{[K-1]},X_{1,\cdots,1,2})\label{fangsub6a}
		\\&\ge\sum _{i=2}^{K-2}H(Z_{[i]},X_{1,\cdots,1,2})+H(W_1,Z_{[K-1]})\notag
		\\&=\sum _{i=2}^{K-2}(H(Z_{[i]},X_{1,\cdots,1,2})+H(W_1,Z_{[i]\cup \{K\}}))-\sum _{i=2}^{K-2}H(W_1,Z_{[i]\cup \{K\}})+H(W_1,Z_{[K-1]})\notag
		\\&\ge\sum _{i=2}^{K-2}(H(W_1,Z_{[i]})+H(Z_{[i]\cup \{K\}},X_{1,\cdots,1,2}))-\sum _{i=2}^{K-2}H(W_1,Z_{[i]\cup \{K\}})+H(W_1,Z_{[K-1]})\label{fangsub6b}
		\\&=\sum _{i=2}^{K-1}H(W_1,Z_{[i]})-\sum _{i=3}^{K-1}H(W_1,Z_{[i]})+\sum _{i=2}^{K-2}H(Z_{[i]\cup \{K\}},X_{1,\cdots,1,2})\label{fangsub6c}
		\\&=H(W_1,Z_{[2]})+\sum _{i=2}^{K-2}H(W_2,Z_{[i]\cup \{K\}},X_{1,\cdots,1,2})\label{fangsub6d}
		\\&\ge H(W_1,Z_{[2]})+(K-3)H(W_2,Z_{[2]},X_{1,\cdots,1,2}),\notag
	\end{align}
\end{subequations}
where (\ref{fangsub6a}) and (\ref{fangsub6d}) hold due to the correctness condition (\ref{decode}), (\ref{fangsub6b}) holds due to the submodular property of the entropy function (\ref{submodular}), and (\ref{fangsub6c}) holds due to the symmetry in user indexing. Thus, we have completed the proofs of all the intermediate inequalities.
\bibliographystyle{IEEEtran}
\bibliography{ref}

\end{document}